\documentclass[11pt,a4paper]{article}
\usepackage{jcappub}
\usepackage{pdflscape}
\usepackage{amsmath}
\usepackage{amssymb}
\usepackage{dcolumn}
\usepackage{bm}
\usepackage{color}
\usepackage{epsfig}
\usepackage{amsfonts}
\usepackage{graphicx}
\usepackage{subfigure}

\begin{document}
		
\title{Non-minimally coupled scalar field cosmology with torsion}
\author[a,b]{Antonella Cid}
\author[c]{Fernando Izaurieta}
\author[d]{Genly Leon}
\author[c]{Perla Medina}
\author[c]{Daniela Narbona}
		
\affiliation[a]{Departamento de F\'isica, Grupo Cosmolog\'ia y Part\'iculas Elementales, Universidad del B\'io-B\'io, Casilla 5-C, Concepci\'on, Chile.}
\affiliation[b]{Observat\'orio Nacional, 20921-400, Rio de Janeiro, RJ, Brasil.}
\affiliation[c]{Departamento de F\'isica, Universidad de Concepci\'on, Casilla 160-C, Concepci\'on, Chile.}
\affiliation[d]{Departamento de Matem\'aticas, Universidad Cat\'olica del Norte, Avenida Angamos 0610, Casilla 1280, Antofagasta, Chile.}

\emailAdd{acidm@ubiobio.cl}
\emailAdd{fizaurie@udec.cl}
\emailAdd{genly.leon@ucn.cl}	
\emailAdd{perlamedina@udec.cl}
\emailAdd{danielanarbona@udec.cl}

\date{\today}

\abstract{In this work we present a generalized Brans-Dicke lagrangian including a non-minimally coupled Gauss-Bonnet term without imposing the vanishing torsion condition. In the resulting field equations, the torsion is closely related to the dynamics of the scalar field, i.e., if non-minimally coupled terms are present in the theory, then the torsion must be present. For the studied lagrangian we analyze the cosmological consequences of an effective torsional fluid and we show that this fluid can be responsible for the current acceleration of the universe. Finally, we perform a detailed dynamical system analysis to describe the qualitative features of the model, we find that accelerated stages are a generic feature of this scenario.}

\maketitle

\section{Introduction}

Scalar-tensor theories or gravity theories with non-minimally coupled scalar fields constitute an alternative to General Relativity, where the existence of additional fields in the gravitational sector may have important consequences in the description of the gravitational interaction \cite{libroFujiiMaeda}. Nowadays, there are a plethora of scalar-tensor theories including the Brans-Dicke theory, where a single dynamical scalar field is added into the gravitational sector \cite{Brans-Dicke}, and the  the Horndeski theory, the most general scenario in a four-dimensional spacetime yielding second order field equations \cite{Horndeski}.

In the context of modern cosmology, scalar-tensor theories are very appealing because the accelerated expanding phases that experience our universe can be described by scalar fields. The first work in scalar-tensor cosmology was developed by Peebles and Dicke \cite{dicke1968scalar}, they consider the Friedmann-Lemaitre-Robertson-Walker (FLRW) metric in the Brans-Dicke theory with the aim of studying the formation of primordial elements in the early universe. Subsequently, La and Steinhardt attempt to solve the graceful exit problem of old inflation with their model of extended inflation \cite{extended}, however, the value obtained for the Brans-Dicke parameter was in tension with observational limits. In spite of this fact, this work promotes scalar-tensor theories as interesting candidates to construct viable and well-motivated cosmological models. Since then, many scalar-tensor cosmological models have been proposed and studied \cite{libroFaraoni}, in particular, scalar-tensor theories have been widely used to model the late-time cosmic acceleration \cite{STDE,STDE2}. 

In General Relativity the spacetime is described by a single rank-2 tensor field, the metric $g_{\mu\nu}$. In the first order formalism, the vierbein (or the metric) and the spin connection (or the affine connection) are independent concepts, so the geometry is not completely determined by the metric and the torsion $T^{a}$ appears, a geometric quality of spacetime which in the second order formalism is fixed by the constraint $T^{a}=0$ \cite{Cartan}. 

The contribution of torsion into gravity theories and particularly into cosmology has been recently studied. In \cite{Poplawski1,Poplawski2} the author considers a fermionic fluid in the matter content, obtaining a cosmological model where the torsion generates accelerated expansion in the early universe. A different approach is developed in \cite{Zanelli-Toloza}, where the authors contemplate the Gauss-Bonnet term coupled to a scalar field and the Einstein-Hilbert term with cosmological constant in the frame of non-vanishing torsion, they find field equations with explicit torsion and present some cosmological scenarios. Furthermore, in \cite{nuestro-otro-paper} it is shown that generic non-minimal couplings of a scalar field and curvature in the Horndeski lagrangian will lead to propagating, non-vanishing torsion. This torsion is ``dark'' for any Yang--Mills interaction, and would interact with fermions only very weakly. For a chronological revision of non-riemannian cosmological models see ref. \cite{nonRiem}.

The current work is an exploration of the feasibility of the idea that torsion could play the role of dark energy in a cosmological setting and, whether or not it could lead to some unrealistic consequences. To study the cosmological solutions of the most general minimal couplings in the full-fledged Horndeski lagrangian as source of torsion would be very difficult, instead we made the choice of the lagrangian (\ref{lagrangian}), which besides to include the most common non-minimal coupling choice in the literature, the Brans-Dicke scenario, it contains the model studied in \cite{Zanelli-Toloza} as a particular case.

Specifically, the aim of this work is to investigate if an effective torsional fluid in the framework of the Horndeski theory with torsion, particularly a generalized Brans-Dicke model, can describe the late-time cosmic evolution without considering any additional scalar field. This paper is organized as follows, in section \ref{uno} we describe a particular case of the Horndeski theory with torsion in the frame of FLRW geometry, in section \ref{dos} we consider a cosmological scenario in order to explain the current acceleration of the universe in terms of an effective torsional fluid. In section \ref{Sect:4} we perform a dynamical system analysis to find the qualitative features of the model at hand. Finally, in section \ref{tres} we present our final remarks.

\section{Torsion, non-minimal couplings and FLRW geometry}\label{uno}

When the null-torsion constraint is relaxed from the beginning in the Horndeski lagrangian, it is possible to prove that non-minimal couplings between the scalar field $\phi$ and the curvature, along with terms involving second order derivatives (such as $\nabla_{\mu}\partial_{\mu}\phi$) in the lagrangian, are sources of torsion \cite{nuestro-otro-paper}. In general, the dynamics of the system strongly departs from the classical riemannian case, and it is possible to have non-vanishing torsion even in the absence of fermionic matter. Let us consider an action principle corresponding to a generalized Brans-Dicke theory with a Gauss-Bonnet term as follows,
\begin{eqnarray}
\label{lagrangian}
S=\int \mathrm{d}x^{4}\sqrt{\left \vert g\right \vert }\left[  \frac{N R}{2\kappa_{4}}-2MX -V+U  \left( R^{2}-4R^{\mu}{}_{\nu}R^{\nu}{}_{\mu}+R^{\mu \nu}{}_{\rho \sigma}R^{\rho \sigma}{}_{\mu \nu}\right)  +\mathcal{L}_{\mathrm{M}}\right],
\end{eqnarray}
where $\mathcal{L}_{\mathrm{M}}$ is the lagrangian for matter, $X=-\frac{1}{2}\partial^{\lambda}\phi\partial_{\lambda}\phi$ and $N$, $M$, $V$ and $U$ are functions of a scalar field $\phi$. $R^{\rho \sigma}{}_{\mu \nu}$ is the Lorentz curvature for the connection $\Gamma_{\mu \nu}^{\lambda}=\mathring{\Gamma}_{\mu \nu}^{\lambda}+K^{\lambda}{}_{\nu \mu}$, where $\mathring{\Gamma}_{\mu \nu}^{\lambda}$ corresponds to the Christoffel connection and $K^{\lambda}{}_{\nu \mu}$ to the contorsion tensor. The Lorentz curvature is related to the standard Riemann $\mathring{R}^{\rho \sigma}{}_{\mu \nu}$ tensor through
\begin{eqnarray}
 R^{\rho \sigma}{}_{\mu \nu}=\mathring{R}^{\rho \sigma}{}_{\mu \nu}+\mathring{\nabla}_{\mu}K^{\rho \sigma}{}_{\nu}-\mathring{\nabla}_{\nu}K^{\rho \sigma}{}_{\mu}+K^{\rho}{}_{\lambda\mu}K^{\lambda \sigma}{}_{\nu}-K^{\rho}{}_{\lambda \nu}K^{\lambda \sigma}{}_{\mu}.
\end{eqnarray}

The lagrangian (\ref{lagrangian}) is a particular case of the Horndeski lagrangian, it corresponds to $F+2W=\frac{1}{4}N(\phi)$, $\kappa_9=-\frac{1}{2}M(\phi)\partial_{\mu}\phi\partial^{\mu}\phi-V(\phi)$ and $\kappa_1=\kappa_3=\kappa_8=0$ in \cite{nuestro-otro-paper} along with a Gauss-Bonnet term non-minimally coupled through the function $U(\phi)$. From the lagrangian (\ref{lagrangian}) it is possible to recover the Brans-Dicke lagrangian \cite{Brans-Dicke} and the scenario studied in \cite{Zanelli-Toloza} by choosing, $N(\phi)=\phi,\ M(\phi)=\frac{\omega}{2\kappa_4\phi},\ V(\phi)=U(\phi)=0$ and $N(\phi)=1,\ M(\phi)=0,\ V(\phi)=\frac{\Lambda}{\kappa_4},\ U(\phi)=\frac{\phi}{4\kappa_4}$, respectively. Moreover, the authors of \cite{KK} study the Kaluza Klein dimensional reduction of the Lovelock Cartan theory in a five-dimensional spacetime. A scalar field is naturally introduced in this scenario through the ansatz for Kaluza Klein compactification in $S^1$. We notice that the dynamics for this model is similar to ours when we consider $N(\phi)=\phi$ and the functions $M(\phi),\ V(\phi),\ U(\phi)$ fixed to constants.

Since the torsion-less condition is not being imposed from the beginning, metricity and parallelism represent different degrees of freedom, therefore, they must be varied independently, \`{a} la Palatini when they are codified by the metric $g_{\mu \nu}$ and the connection $\Gamma_{\mu \nu}^{\lambda}$, or \`{a} la Cartan when they are codified by the vierbein $e^{a}$ and the spin connection $\omega^{ab}$. In this context, the equations of motion corresponding to the lagrangian \eqref{lagrangian} are given by
\begin{eqnarray}
\mathcal{E}_{\mu \nu}&=& N\left(  R_{\mu \nu}-\frac{1}{2}g_{\mu \nu}R\right)  +\kappa_{4}\left(-2MX g_{\mu \nu}+2V g_{\mu \nu}-2M\partial_{\mu}\phi \partial_{\nu}\phi-\mathcal{T}_{\mu \nu}\right)=0,\label{E_metrica}\\
\mathcal{E}_{\mu \nu \lambda}  & =& N\left(T_{\lambda \mu \nu}+T^{\rho}{}_{\rho \mu}g_{\lambda \nu}-T^{\rho}{}_{\rho \nu}g_{\lambda \mu}\right)  +\left(  \frac{\partial N}{\partial \phi}+4\kappa_{4}\frac{\partial U}{\partial \phi}R\right) \left(  g_{\lambda \mu}\partial_{\nu}\phi-g_{\lambda \nu}\partial_{\mu}\phi \right)-\kappa_{4}\sigma^{\lambda}{}_{\mu \nu}      \nonumber \\
&& + 8\kappa_{4}\frac{\partial U}{\partial \phi}\left( \partial_{\rho}\phi \left(  R^{\rho}{}_{\mu}g_{\lambda \nu}-R^{\rho}{}_{\nu}g_{\lambda \mu}-R^{\rho \lambda}{}_{\mu \nu}\right)  +\partial_{\mu}\phi R_{\lambda \nu}-\partial_{\nu}\phi R_{\lambda \mu}\right)   =0,\label{E_conexion}\\
 \mathcal{E}&= & \frac{1}{2\kappa_{4}}\frac{\partial N}{\partial \phi}R- \frac{\partial M}{\partial \phi}X-\frac{\partial V}{\partial \phi} +M\nabla_{\mu}\partial^{\mu}\phi+M{\partial^{\nu}\phi T^{\mu}{}_{\nu\mu}}\nonumber\\
&&+\frac{\partial U}{\partial \phi}\left(  R^{2}-4R^{\mu}{}_{\nu}R^{\nu}{}_{\mu}+R^{\mu \nu}{}_{\rho \sigma}R^{\rho \sigma}{}_{\mu \nu}\right) =0,\label{E_escalar}
\end{eqnarray}
where $\mathcal{T}_{\mu \nu}$ is the energy-momentum tensor associated to $\mathcal{L}_{\mathrm{M}}$, $\sigma^{\lambda}{}_{\mu \nu}$ is the spin tensor associated to $\mathcal{L}_{\mathrm{M}}$, and $T^{\lambda}{}_{\mu \nu}$ is the torsion.

The equation~(\ref{E_conexion}) implies that the torsion depends on the derivatives of the scalar field, and it does not vanish even when $\sigma^{\lambda}{}_{\mu \nu}=0$. The terms generating this feature are precisely the non-minimal couplings. This behavior stands in strong contrast with the standard minimally coupled Einstein--Cartan case, where only the spin tensor associated to fermions can give rise to torsion, in such a way that it can not propagate in vacuum.

Another interesting feature is that setting $T^{\lambda}{}_{\mu \nu}=0$ into equations  \eqref{E_metrica}--\eqref{E_escalar}  does not lead to the standard expressions we would get for the standard torsion-less case with the Christoffel connection. Instead, imposing $T^{\lambda}{}_{\mu \nu}=0$ on the equations of motion necessarily freezes the scalar field, leading to $\partial_{\mu}\phi=0$.

This behavior reveals that the torsional and torsion-less cases correspond to different dynamical systems when non-minimal couplings are present (see reference \cite{nuestro-otro-paper} for a treatment on the full Horndeski lagrangian\footnote{In \cite{nuestro-otro-paper} the results are presented in the first order formalism, in the language of differential forms. The Lagrange multiplier $\lambda^{\lambda \mu \nu}$ in this article and the 2-form Lagrange multiplier $\Lambda^{a}$ of \cite{nuestro-otro-paper} are related through $\frac{1}{2}e^{a}{}_{\beta}\lambda^{\beta}{}_{\mu \nu}\mathrm{d}x^{\mu}\wedge \mathrm{d}x^{\nu}=-\ast \Lambda^{a}$.}). The appropriate procedure to recover the torsion-less case is to include a Lagrange multiplier constraint into the action (\ref{lagrangian}),
\begin{eqnarray}
\bar{S}=S+\int \mathrm{d}x^{4}\sqrt{\left \vert g\right \vert }\frac{1}{2}\lambda^{\lambda \mu \nu}T_{\lambda \mu \nu}.
\end{eqnarray}
In our work, the equations of motion for the torsion-less constrained case are recovered:
\begin{eqnarray*}
 \overline{\mathcal{E}}^{\mu \nu}  &  =&\mathcal{E}^{\mu \nu}+\frac{1}{2}\nabla_{\lambda}\left(  \lambda^{\mu \nu \lambda}+\lambda^{\nu \mu \lambda}\right)  =0,\\
 \overline{\mathcal{E}}^{\mu \nu \lambda}  &  =&\mathcal{E}^{\mu \nu \lambda}-\frac{1}{2}\left(  \lambda^{\mu \nu \lambda}-\lambda^{\nu \mu \lambda}\right)=0,\\
 \overline{\mathcal{E}}  &  =&\mathcal{E}=0,\\
 T_{\lambda \mu \nu}  &  =&0,
\end{eqnarray*}
where the solution to this system corresponds to the classical riemannian scenario,
\begin{eqnarray}
 \left.  \mathcal{E}^{\mu \nu}+\nabla_{\lambda}\left(  \mathcal{E}^{\lambda\mu \nu}+\mathcal{E}^{\lambda \nu \mu}\right)  \right \vert _{T_{\alpha \beta \gamma}=0}  &  =0,\label{riem1}\\
 \left.  \mathcal{E}\right \vert _{T_{\alpha \beta \gamma}=0}  &  =0.\label{riem2}
\end{eqnarray}

It is worth to notice that the field equations (\ref{riem1}) and (\ref{riem2}), considering the definitions (\ref{E_metrica})--(\ref{E_escalar}), are reduced to the field equations of the Brans-Dicke theory in the Jordan frame, presented for instance in reference \cite{Pedro2}, corresponding to $N(\phi)=\phi,\ M(\phi)=\frac{\omega}{2\phi}$ and $U(\phi)=0$ in our scenario.

The role of torsion in cosmology has been studied in the frame of standard Einstein--Cartan geometry without scalar fields \cite{Poplawski1,Poplawski2}. In this context, torsion can play an important role only inside of a very dense fermionic plasma, as in the Big Bounce model presented in \cite{Poplawski2}. Inside of standard fermionic matter at usual densities, the effects of torsion are extremely small, see for instance chapter 8 of reference \cite{Supergravity-VanProeyen}. Torsion generated by non-minimal couplings in the context of cosmology was presented in \cite{Zanelli-Toloza}, but only for the Gauss-Bonnet coupling. In \cite{Karpathopoulos:2017arc} it was investigated the cosmology of a higher-order modified teleparallel theory by means of  analytical cosmological solutions. In particular, there were determined forms of the unknown potential which drives the scalar field such that the field equations form a Liouville integrable system. The conservations laws were determined using the Cartan symmetries.

On the other hand, in the standard cosmological scenario, the requirement of spatial homogeneity and isotropy is reflected on the condition
\begin{equation*}
\pounds _{\vec{\zeta}}\ g_{\mu \nu}=0,
\end{equation*}
for the vector fields $\vec{\zeta}$ corresponding to spatial translations and rotations, leading to the Friedmann-Lemaitre-Robertson-Walker metric,
\begin{equation*}
ds^2=-c^2dt^2+a^2(t)\left(\frac{dr^2}{1-kr^2}+r^2d\theta^2+r^2\sin^2\theta\  d\phi^2\right),
\end{equation*}
depending only in the  scale factor $a\left(  t\right)$, where $(t,r,\theta,\phi)$ are the coordinates in the co-moving frame. From here on we will consider $\kappa_4=c=1$. A non-vanishing torsion implies new independent degrees of freedom for the geometry, which must also satisfy spatial homogeneity and isotropy in a cosmological setup. These degrees of freedom are codified in the contorsion tensor,
\begin{equation*}
K^{\lambda}{}_{\nu \mu}=\Gamma_{\mu \nu}^{\lambda}-\mathring{\Gamma}_{\mu \nu}^{\lambda},
\end{equation*}
and therefore we must require
\begin{equation*}
\pounds _{\vec{\zeta}}\ K^{\lambda}{}_{\mu \nu}=0,
\end{equation*}
for vector fields $\vec{\zeta}$ corresponding to spatial translations and rotations. This leads to a \textquotedblleft FLRW contorsion\textquotedblright \ parametrized in four dimensions as
\begin{equation*}
 K_{\mu \nu \lambda}=\left(  g_{\mu \lambda}g_{\nu \rho}-g_{\mu \rho}g_{\nu \lambda}\right)  h^{\rho}+\sqrt{\left \vert g\right \vert }\epsilon_{\mu \nu \lambda \rho}f^{\rho},
\end{equation*}
or a FLRW torsion given by
\begin{equation*}
T_{\lambda \mu \nu}=\left(  g_{\mu \lambda}g_{\nu \rho}-g_{\mu \rho}g_{\nu \lambda}\right)  h^{\rho}-2\sqrt{\left \vert g\right \vert }\epsilon_{\lambda \mu \nu\rho}f^{\rho},
\end{equation*}
where $h^{\rho}$ and $f^{\rho}$ are two vectors that in the co-moving frame take the form
\begin{equation*}
 h_{0} =h\left(t\right),\quad
 h_{i} =0,\quad
 f_{0} =f\left(t\right)\quad \textrm{and}\quad
 f_{i} =0,
\end{equation*}
with $i=1,2,3.$

In this sense, the cosmological evolution is characterized by the temporal evolution of three functions $a\left(t\right)$, $h\left(t\right)  $ and $f\left(t\right)$, instead of only $a\left( t\right)  $ as in the classical riemmanian case. For the sake of simplicity, from here on, we will suppose only classical spin-less matter with vanishing spin tensor
\begin{equation*}
 \sigma_{\lambda \mu \nu}=0.
\end{equation*}
It means that we will consider torsion produced only by the scalar field $\phi$. 

Finally, by considering the energy-momentum tensor for a perfect fluid, $\mathcal{T}_{\mu \nu}=\left(p+\rho\right)u_{\mu}u_{\nu}+pg_{\mu \nu}$ (where $\rho$ is the energy density, $p$ is the pressure and $u^{\mu}$ the four-velocity of an observer co-moving with the fluid), the generalized equations for the FLRW geometry are given by
\begin{eqnarray}
3N\left(  \left(  H+h\right)  ^{2}+\frac{k}{a^{2}}-f^{2}\right)  -\left(  \frac{M\dot{\phi}^{2}}{2}+V\right)   &=&\rho,\label{E1}\\
N\left(2(\dot{H}+\dot{h})+(3H+h)(H+h)+\frac{k}{a^{2}}-f^{2}\right) + \frac{M\dot {\phi}^{2}}{2} -V &=&-     p, \label{E2}\\	
Nh-\dot{\phi}\left(  \frac{1}{2}\frac{\partial N}{\partial \phi}+4\frac{\partial U}{\partial \phi}\left(  \left(  H+h\right)  ^{2}+\frac{k}{a^{2}}-f^{2}\right)
\right)  &=&0,\label{E3}\\
f\left(N-8\frac{\partial U}{\partial \phi}\dot{\phi}\left(  H+h\right)\right)   &=&0,\label{E4}
\end{eqnarray}
where we have considered the Hubble expansion rate $H=\frac{\dot{a}}{a}$ and dots denote derivatives with respect to the cosmic time $t$. In the same way, the equation of motion for the scalar field is given by
\begin{eqnarray}
\label{E5}
&  3\frac{\partial N}{\partial \phi}\left (  \dot{H}+\dot{h}+H(H+h)  + (H+h)^{2}+\frac{k}{a^{2}}-f^{2}  \right )  -\frac{1}{2}\frac{\partial M}{\partial \phi}\dot{\phi}^{2}-\frac{\partial V}{\partial \phi}-M\left ( \ddot{\phi}+3\dot{\phi}H \right ) \nonumber\\
& +24\frac{\partial U}{\partial \phi}\left (  \left(  \dot{H}+\dot{h}+H(H+h)\right)  \left(  (H+h)^{2}+\frac{k}{a^{2}}-f^{2}\right)-2f(H+h)(\dot{f}+Hf)\right )  =0,
\end{eqnarray}
and the conservation equation for the energy-momentum tensor become
\begin{equation}
\dot{\rho}+3H(\rho+p)=0,\label{TT3}
\end{equation}
where the equation of state will be provided by a barotropic fluid where $p=\omega\rho$, for a constant state parameter $\omega$. Notice that equation \eqref{TT3} is obtained by combining equations \eqref{E1}--\eqref{E5}, and in this sense it is not an independent equation. 

\section{Cosmological Scenario}\label{dos}

In cosmology it is common to consider effective fluids in the study of scalar-tensor theories of gravity \cite{Starobinsky}, given that in our scenario the torsion is closely related to the scalar field, we define an effective fluid associated to the torsion by
\begin{eqnarray}
\label{rhoT}
\rho_T&=&\frac{M\dot{\phi}^2}{2}+V+3\left(H^2+\frac{k}{a^2}\right)(1-N)-3N\left(2Hh+h^2-f^2\right),\\
\label{pT}
p_T&=&\frac{M\dot{\phi}^2}{2}-V-\left(3H^2+\frac{k}{a^2}+2\dot{H}\right)\left(1-N\right)+N\left(2\dot{h}+4Hh+h^2-f^2\right),
\end{eqnarray} 
in such a way that the field equations \eqref{E1}, \eqref{E2}, \eqref{E5} and \eqref{TT3}  can be cast in the standard form for a universe filled with two non-interacting fluids, $\rho$ and $\rho_T$, as
\begin{eqnarray}
3\left(H^2+\frac{k}{a^2}\right)&=&\rho+\rho_T,\label{T1}\\
2\dot{H}+3H^2+\frac{k}{a^2}&=&-p-p_T,\label{T2}\\
\dot{\rho_T}+3H(\rho_T+p_T)&=&0\label{T4},\\
\dot{\rho}+3H(\rho+p)&=&0.\label{T3}
\end{eqnarray}
Furthermore, equations \eqref{E3} and \eqref{E4} have to be satisfied. 

Notice that we can recover the expressions for the effective density and pressure defined in \cite{Zanelli-Toloza} by assuming $N(\phi)=1$, $M(\phi)=0$ and $V(\phi)=\Lambda$ into equations  \eqref{rhoT} and \eqref{pT}, where we have generalized the cosmological constant to the potential of the scalar field $\phi$, including it as part of the total energy-momentum tensor instead of the gravitational sector as in \cite{Zanelli-Toloza}.

On the other hand, following reference \cite{Zanelli-Toloza}, we find conditions to analytically solve the system of equations (\ref{E1})--(\ref{E5}), where an effective torsional fluid defined through (\ref{rhoT}) and (\ref{pT}), drives the acceleration in a cosmological setup. In order to do this we rewrite the set of equations (\ref{E1})--(\ref{E5}) in the following way:
\begin{eqnarray}
\label{Eq1}
Z^{2}+W-\frac{V}{3N} - \frac{1}{3N}\left(\rho+\frac{1}{2}M\dot{\phi}^{2}\right)&=&0,\\
\label{Eq2}
2\dot{Z}+2HZ+Z^{2}+W-\frac{V}{N}+\frac{1}{N}\left(p+\frac{1}{2}M\dot{\phi}^{2}\right)&=&0,\\
\label{Eq3}
-Nh+4U'\dot{\phi}\left(Z^{2}+W\right) +\frac{1}{2}N'\dot{\phi}&=&0,\\
\label{Eq4}
f\left(N-8U'\dot{\phi}Z\right)&=& 0,\\
\label{Eq5}
\left(\dot{Z}+HZ+Z^{2}+W\right)N'-\dfrac{1}{3}\left(\frac{1}{2}M'\dot{\phi}^{2}+V'+M\left(\ddot{\phi} +3H\dot{\phi}\right)\right)&& \nonumber\\
+8U'\left(\left(Z^{2}+W\right)\left(\dot{Z}+HZ\right)-2fZ\left(\dot{f}+Hf\right)\right)&=& 0,
\end{eqnarray}
where prime denotes derivative with respect to $\phi$ and we have defined:
\begin{eqnarray}
\label{Z0}
Z & = & H+h,\\
\label{W0}
W & = & \frac{k}{a^{2}}-f^{2},
\end{eqnarray}
Notice that the cosmological scenario presented in \cite{Zanelli-Toloza} is recovered by considering $N(\phi)=1$, $M(\phi)=0$, $V(\phi)=\Lambda$, $U(\phi)=\frac{\phi}{4}$ into equations \eqref{Eq1}--\eqref{Eq5}. In the following subsections we present analytical solutions to the set of equations  \eqref{Eq1}--\eqref{Eq5}.

\subsection{Solution for $f\neq0$ with pressureless matter}
\label{cosmo}

In the late-time universe the energy-momentum tensor must consider baryonic and dark matter, both of which behave as pressureless matter ($p=0$) at cosmological scales.

By assuming $f\neq0$, we can replace $U'$ from equation \eqref{Eq4} into equation (\ref{Eq3}) to get
\begin{eqnarray}
\label{W}
W = Z^2-2HZ-\frac{N'}{N}Z\dot{\phi},
\end{eqnarray}
then, by considering $p=0$, replacing (\ref{W}) into equation (\ref{Eq2}) and rewriting $\dot{\phi}$ in terms of $U'$ through equation (\ref{Eq4}) we obtain
\begin{eqnarray}
\label{key}
2(\dot{Z}+Z^{2})=\frac{N'}{8U'}+\frac{V}{N} -\frac{MN}{128U'^2Z^2},
\end{eqnarray}
where we notice that by setting the following conditions:
\begin{eqnarray}
\label{conditions}
M(\phi)=0,\quad V(\phi)=\beta N(\phi),\quad U'(\phi)=\alpha N'(\phi),
\end{eqnarray}
we are able to find an analytical solution for $\alpha$ and $\beta$ constants. This scenario corresponds to a generalized Brans-Dicke theory without the kinetic term and including a non-minimally coupled Gauss-Bonnet term. By replacing \eqref{conditions} into equation \eqref{key} we get:
\begin{equation}
\label{Z}
Z(\tau)=Z_{e}\ \textrm{tanh} (\tau),
\end{equation}
where $Z_{e}=\sqrt{\frac{\beta}{2}+\frac{1}{16\alpha}}$, $\tau=Z_{e}(t-t_i)$ is a rescaled dimensionless time, $t_i$ is an integration constant and we have imposed the condition $\frac{1}{\alpha}+8 \beta>0$. Since $Z_{e}\geq 0$, $Z$ is positive for $t>t_i$, and negative for $t<t_i$. Besides, by integrating equation (\ref{Eq4}) we have
\begin{equation}
\label{N}
N(\tau)=N_*\ \textrm{sinh}(\tau)^{2-3x},
\end{equation}
where $N_*$ is an integration constant, and for convenience we define the dimensionless parameter $x=\beta Z_{e}^{-2}/3$. Also,  by integrating \eqref{T3} for $p=0$ we get $\rho=\rho_{m0}\left(\frac{a_0}{a}\right)^3$, where the subscript 0 indicates current values. By using this last result into equation (\ref{Eq1}) and considering equation \eqref{W} we obtain
\begin{eqnarray}
\label{da}
N\left(6Z^{2}-3Z\left(2\frac{\dot{a}}{a}+\frac{\dot{N}}{N}\right)-\beta \right) = \rho_{m0}\left(\frac{a_0}{a}\right)^3.
\end{eqnarray}
Finally, by inserting equations (\ref{Z}) and (\ref{N}) into equation (\ref{da}) we are able to integrate, obtaining the following dimensionless scale factor
\begin{equation}
\label{scalefactor}
\bar{a}(\tau)=\frac{a}{a_0}=\bar{a}_0\ \textrm{cosh}(\tau)\ \textrm{sinh}(\tau)^{-1+x}\left(1-y\ \textrm{tanh}(\tau)\right)^{1/3},
\end{equation}
where we define the dimensionless parameter $y=\frac{\rho_{m0}}{2\bar{a}_0^{3}Z_{e}^2N_*}>0$ and $\bar{a}_0$ is a positive defined integration constant. Notice that the behavior of the scale factor is independent of the exact functional form of $N(\phi)$ when $V(\phi)$ and $U(\phi)$ are defined by \eqref{conditions}.

In the same way, we obtain $h(\tau)$ and $f(\tau)$ from equations \eqref{Z0} and \eqref{W0}, respectively:
\begin{eqnarray}
h(\tau)&=&Z_{e}\left((1-x)\ \textrm{coth}(\tau)+\frac{y\ \textrm{sech}^2(\tau)}{3\left(1-y\ \textrm{tanh}(\tau)\right)}\right),\\
f(\tau)&=&\pm\sqrt{\frac{k\ \textrm{tanh}^2(\tau)\left(1-y\ \textrm{tanh}(\tau)\right)^{-2/3}}{\bar{a}_0^{2}\ \textrm{sinh}(\tau)^{2x}}-\frac{\text{Z}_{e}^2 \left(x-Y\ \text{tanh}(\tau )\right)}{1- y\ \text{tanh}(\tau)}},
\end{eqnarray}
where $Y=y \left(-1+ x+\text{sech}(\tau)^2/3\right)+ \text{tanh}(\tau )$.

Without loss of generality, we can assume  $t_i=0$ in order to show the behavior of the scale factor depending on the parameters $x$ and $y$ as we see in table \ref{table1}.
\begin{table}[h!]
\centering
\scalebox{0.750}{
\begin{tabular}{|c|c|l|}
\hline
Label&Conditions & 
\multicolumn{1}{|c|}{Behavior of the dimensionless scale factor} \\\hline
$\it{S_1}$&  $1<x<2$  and $y<1$ & $\bar{a}(0)=0$, then it expands decelerated and subsequently accelerated\\\hline
$\it{S_2}$& $x>2$  and $y<1$ & $\bar{a}(0)=0$, then it accelerated expands\\ \hline
$\it{S_3}$&  $x>1$  and $y>1$ & $\bar{a}(0)=0$, then it expands to a maximum, subsequently it contracts and eventually collapses\\\hline
$\it{S_4}$& $0<x<1$  and $y<1$ &$\bar{a}(0)\rightarrow\infty$, then it contracts to a minimum and subsequently it accelerated expands \\\hline
$\it{S_5}$&    $x<1$  and $y>1$ & $\bar{a}(0)\rightarrow\infty$ then it contracts and eventually collapses \\\hline
$\it{S_6}$&  $x<0$  and $y<1$ & $\bar{a}(0)\rightarrow\infty$, then it contracts and subsequently it becomes asymptotically zero \\\hline
$\it{S_7}$& $x=1$  and $y<1$ & $\bar{a}(0)=\bar{a}_0$, then it accelerated expands\\ \hline
$\it{S_8}$& $x=1$  and $y>1$ & $\bar{a}(0)=\bar{a}_0$, then it contracts and eventually collapses \\ \hline
$\it{S_9}$&  $x=0$  and $y<1$ & $\bar{a}(0)\rightarrow\infty$, then it contracts and subsequently it becomes asymptotically constant \\ \hline
$\it{S_{10}}$&   $x=0$  and $y>1$ & $\bar{a}(0)\rightarrow\infty$, then it contracts and eventually collapses \\ \hline
\end{tabular}}
\caption{\label{table1} Summary of the different behaviors of the scale factor depending on the relations among the parameters $x$ and $y$. Note that the parameter $y$ is positive defined.}
\end{table}

\begin{figure}[ht!]
\centering
\includegraphics[width=0.5\textwidth]{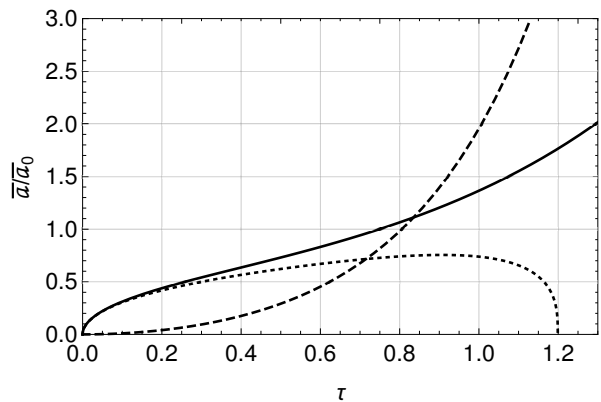}
\caption{\label{fig1}The behavior of the dimensionless scale factor for $\it{S_1}$, $\it{S_2}$ and $\it{S_3}$ in table \ref{table1}. Depending on the value of the parameters we have different scenarios: a transition from decelerated expansion to accelerated expansion for $(x,y)=(1.5,0.6)$ (solid line), accelerated expansion for $(x,y)=(3,0.3)$ (dashed line) and a collapsing scenario for $(x,y)=(1.5,1.2)$ (dotted line).}
\end{figure}

From the cosmological point of view, an interesting behavior is associated with $x>1$, where the scale factor starts with a null value at $\tau=0$ and subsequently expands. In figure \ref{fig1} we show different possible behaviors for $x>1$, depending on the values for the parameters. Among these models, the most interesting scenario is the one with a transition from decelerated expansion to  accelerated expansion, the case $\it{S_1}$ in table \ref{fig1}.

For the scenario $\it{S_1}$, we obtain the Hubble expansion rate and the deceleration parameter $q=-\frac{\ddot{a}a}{\dot{a}^2}$ by deriving (\ref{scalefactor}),
\begin{eqnarray}
H&=&Z_{e} \left(\textrm{tanh} (\tau )+\left(x-1\right)\textrm{coth} (\tau )-\frac{y\  \textrm{sech}^2(\tau )}{3-3y\ \textrm{tanh}(\tau)}\right),\label{H}\\
q&=&-1-\frac{Z_{e}^2}{3H^2}  \left(2\ \textrm{sech}^2(\tau )-3\left(x-1\right)\ \textrm{csch}^2(\tau )+\frac{\left(1-y^2\right)\textrm{sech}^2(\tau)}{\left(1-y\ \textrm{tanh}(\tau )\right)^2}\right).\label{q}
\end{eqnarray}
From (\ref{q}) we notice that in the limit $\tau\rightarrow0$ we have $q\rightarrow\frac{2-x}{-1+x}$, a positive value for $1<x<2$, meanwhile in the limit $\tau\rightarrow \infty$ we get $q\rightarrow-1$, then we can see that for $1<x<2$ necessarily exist a sign-transition going from decelerated expansion to accelerated expansion, scenario favored by current data.

\begin{figure}[ht!]
\centering
\includegraphics[width=0.47\textwidth]{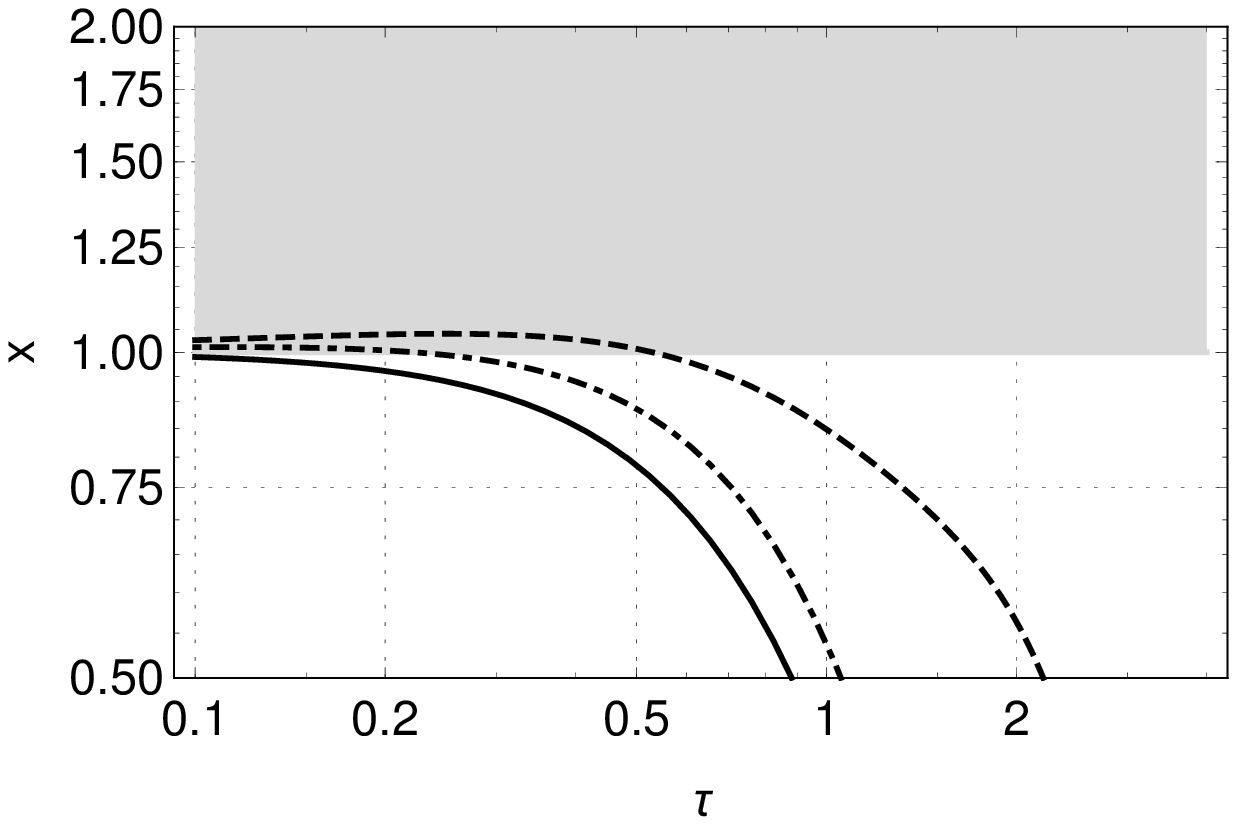}
\includegraphics[width=0.46\textwidth]{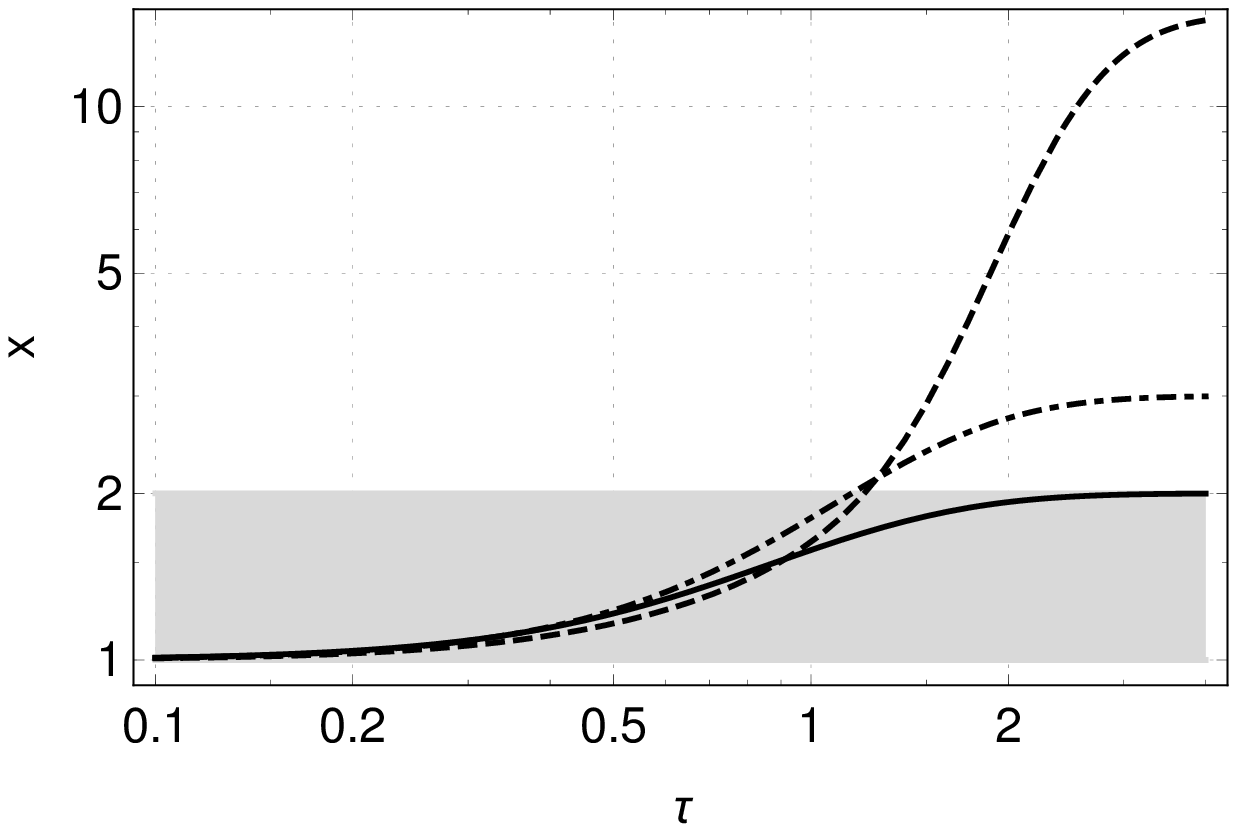}
\caption{\label{fig2}The curves show the lower limit of $x$, from inequality (\ref{condition}) in the left panel and from inequality (\ref{condition2}) in the right panel, for different choices of the parameter $y$. The solid, dot-dashed and dashed lines correspond to $y=0,\ 0.6,\ 0.95$, respectively. The shaded regions represent the allowed range of the parameter $x$ in order to have a sign-transition in the deceleration parameter.}
\end{figure}

On the other hand, if we want to impose the conditions $H(t)>0$ and $\dot{H}(t)<0$, as in the standard cosmological scenario \cite{wein}, we find that the following constraints 
\begin{eqnarray}
\label{condition}
\frac{ (1-x)(y\text{coth}(\tau )- \text{coth}^2(\tau ))-y \text{csch}(\tau ) \text{sech}(\tau )/3}{ y- \text{coth}(\tau )}&<&\text{tanh}(\tau ),\\
\label{condition2}
3 x+2 \text{sech}^2(\tau )+\frac{1-2 y \text{sinh}(\tau )(\text{cosh}(\tau )-y \text{sinh}(\tau ))}{(\text{cosh}(\tau )-y \text{sinh}(\tau ))^2}&>&6,
\end{eqnarray}
must be satisfied, respectively. Nevertheless, in figure \ref{fig2} we show some numerical examples indicating that these conditions are inconsistent with $1<x<2$ and $0<y<1$.
In figure \ref{fig2} the curves represent the lower limit of the parameter $x$ satisfying inequality \eqref{condition} (left panel) and inequality \eqref{condition2} (right panel) for $0<y<1$, the shaded regions represent $1<x<2$. We notice that during the evolution the condition $\dot{H}<0$ becomes inconsistent for $\tau>1$, then we conclude that if a transition from decelerated to accelerated expansion exists, necessarily occurs a period with $\dot{H}\geq0$. This behavior of the Hubble expansion rate has been considered in the literature before, in order to describe the transition from a static state to an inflationary phase in Emergent Universe scenarios \cite{Pedro}, where the regime $\dot{H}\geq0$ is transitory. By contrast, in our scheme, this behavior corresponds to the final stage of evolution.

\begin{figure}[ht!]
\centering
\includegraphics[width=0.47\textwidth]{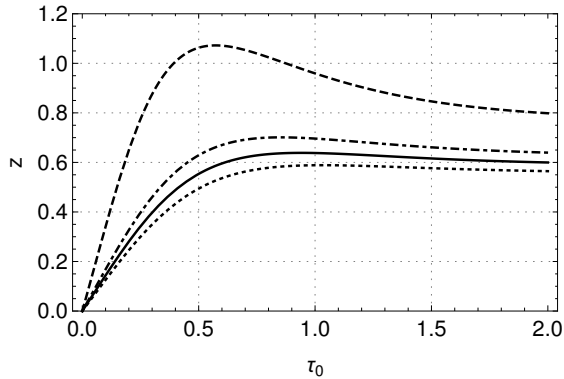}
\caption{\label{fig6}The figure shows the time today $\tau_0$ in terms of the parameter $z$ for a given $x$ and $y$. The dotted, solid, dot-dashed and dashed lines are for $y=0.5$ and $x=1.8,\ 1.7,\ 1.6,\ 1.3$, respectively.} 
\end{figure}

By convenience, we choose to measure the parameter $Z_e$ in terms of the Hubble parameter $H_0$ as $Z_e=zH_0$, where $z$ is a dimensionless parameter. On the other hand, by imposing the value of the expansion rate today to be $H_0$, we find a constraint among the parameters $x$, $y$ and $z$, that is, there exist a maximum value for the parameter $z$ over which the consistency condition $H=H_0$ can not be reached at any time. For instance, in figure  \ref{fig6} we see that for $x=1.8$ and $y=0.5$, the maximum possible value of $z$ is around 0.6.

\begin{figure}[ht!]
\centering
\includegraphics[width=0.47\textwidth]{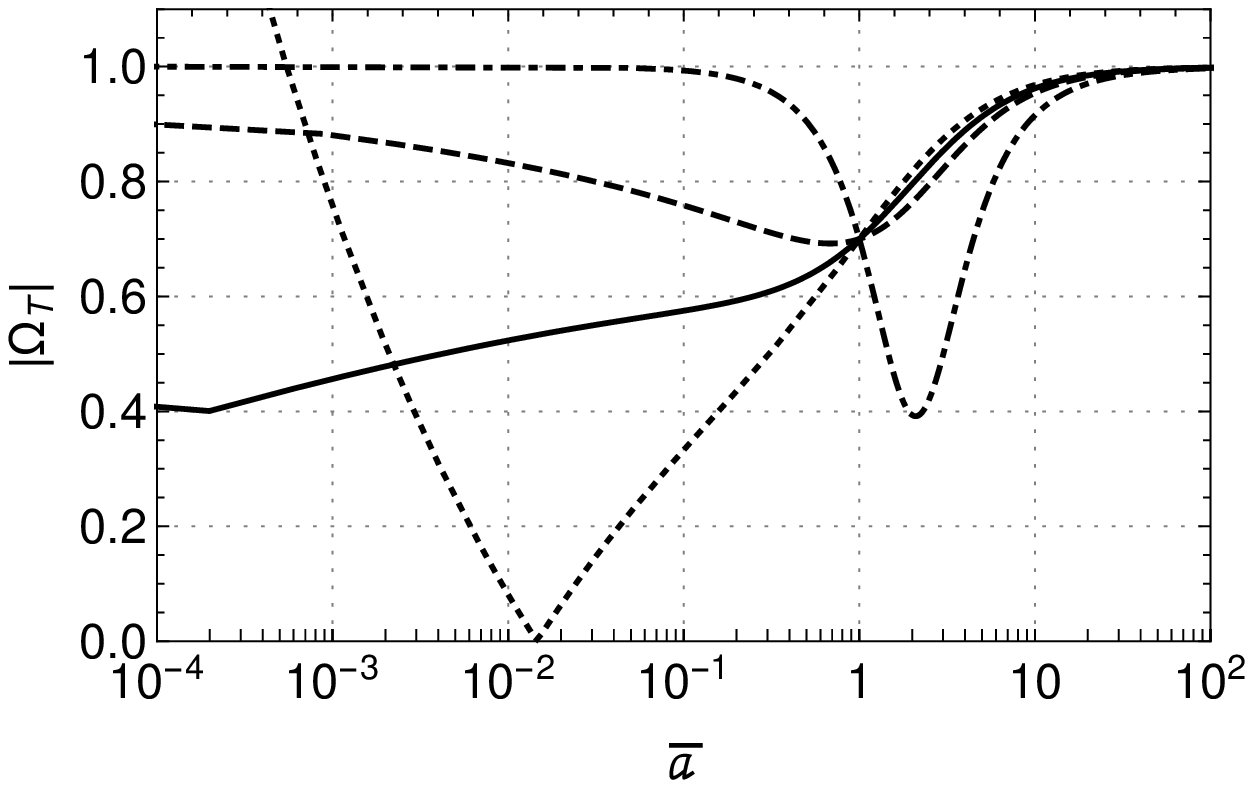}
\includegraphics[width=0.47\textwidth]{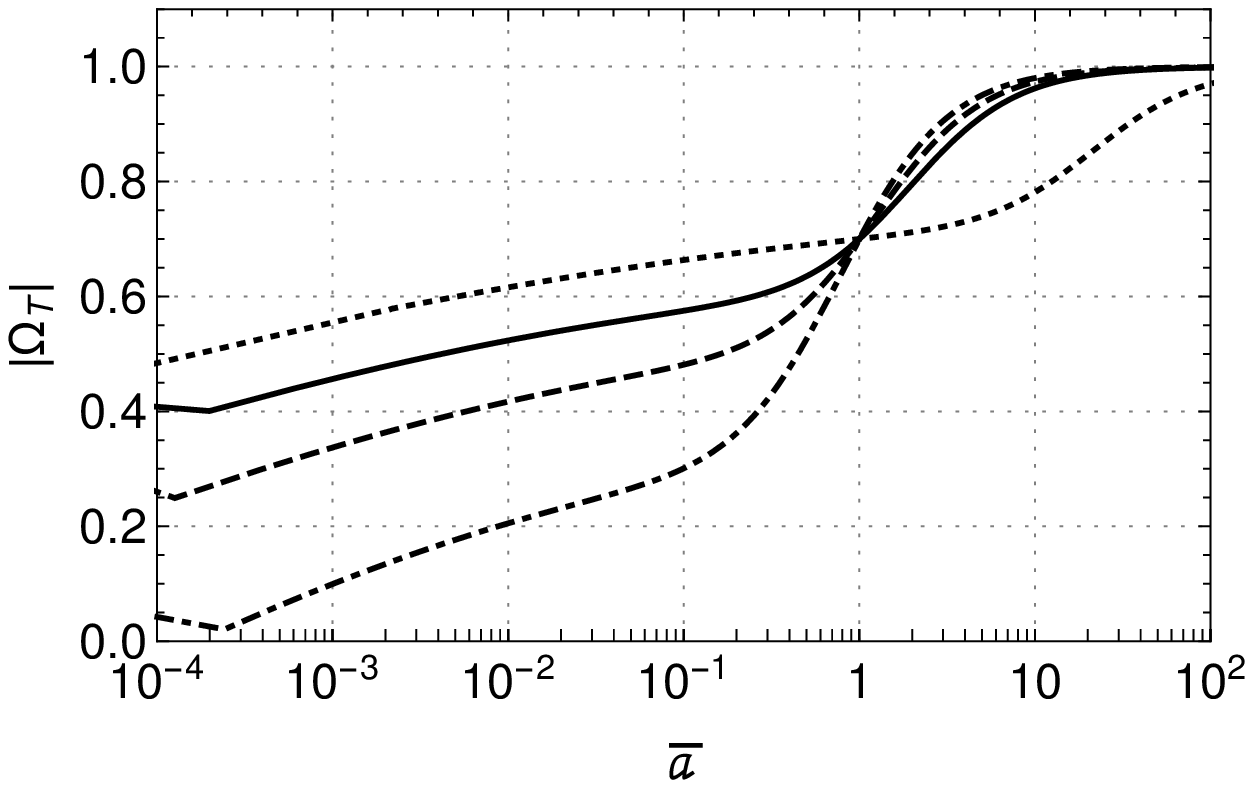}
\caption{\label{fig5}The figures show the behavior of the absolute value of $\Omega_T$. The left panel is for $y=0.5$ and $z=0.4$, the dotted, solid, dashed and dot-dashed lines stand for $x=1.8,\ 1.7,\ 1.6,\ 1.3$, respectively. In the right panel we have $x=1.7$, $y=0.5$ and $z=0.1,\ 0.4,\ 0.48,\ 0.55$ for dotted, solid, dashed and dot-dashed lines, respectively.} 
\end{figure}

\begin{figure}[ht!]
\centering
\includegraphics[width=0.5\textwidth]{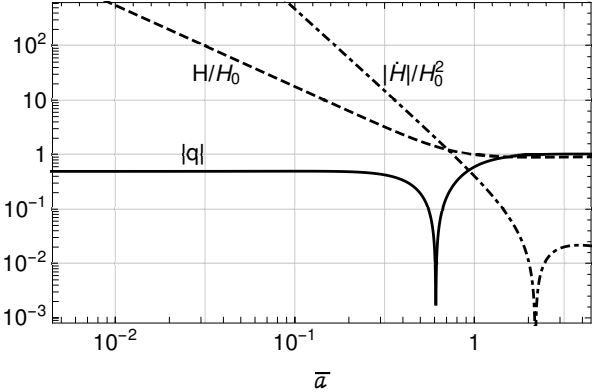}
\caption{\label{figa} The figure shows the behavior of $H/H_0$ (dashed line), {the absolute value of $q$} (solid line) and {the absolute value of $\dot{H}/H_0^2$} (dot-dashed line), {for $x=1.67$, $y=0.24$ and $z=0.56$}.}
\end{figure}

\begin{figure}[ht!]
\centering
\includegraphics[width=0.47\textwidth]{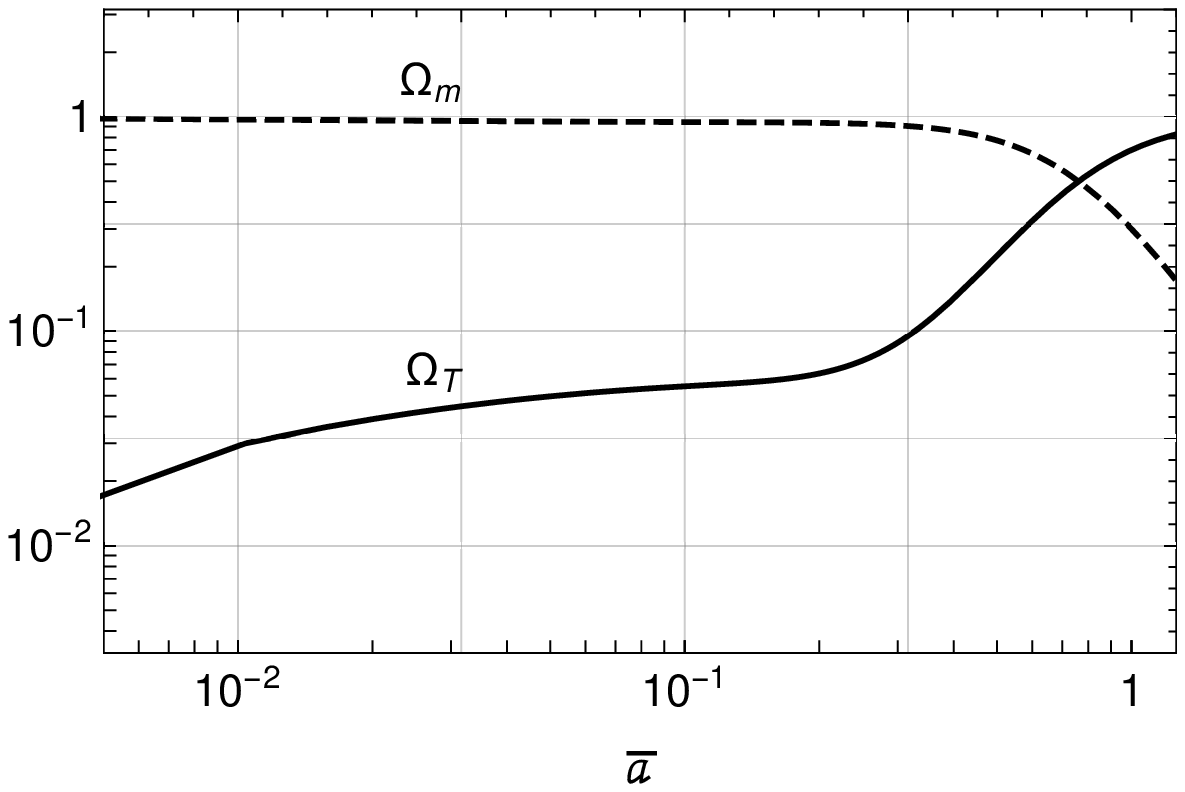}
\includegraphics[width=0.5\textwidth]{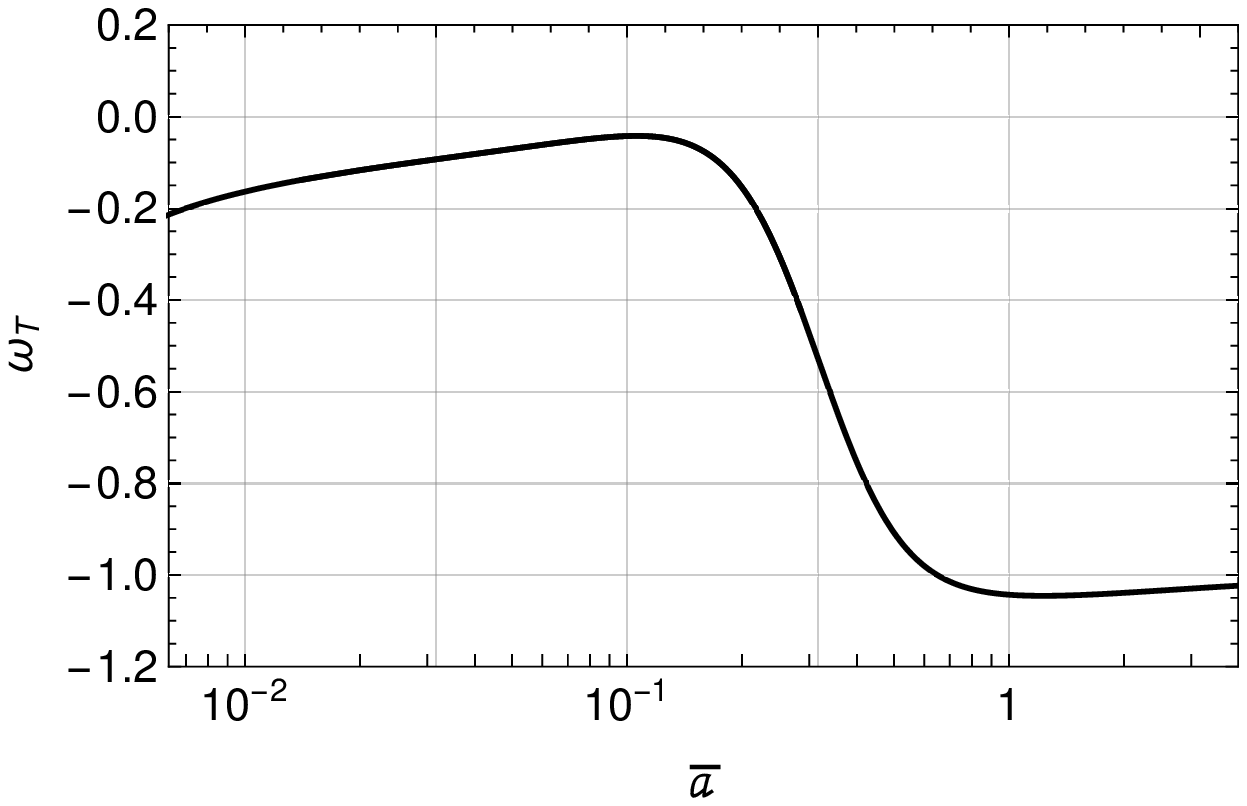}
\caption{\label{figb} The left panel shows the behavior of the density parameters, $\Omega_T$ (solid line) and $\Omega_m$ (dashed line). The right panel shows the evolution of the state parameter for the effective torsional fluid. We fixed the parameters to $x=1.67$, $y=0.24$, $z=0.56$, $k=0$ and $\Omega_{m0}=0.3$.}
\end{figure}

In figures \ref{fig5}--\ref{figb} we have used the dimensionless scale factor $\bar{a}$ to give account of the evolution in our cosmological scenario, where $\bar{a}$ is related to the physical redshift $\bar{z}$ by $\bar{a}^{-1}=1+\bar{z}$. Besides, given that we are interested in the periods of matter and dark energy domination, we focus on the range $\bar{z}<200$ or $\bar{a}>5\times10^{-3}$, when the radiation contribution in the standard cosmological scenario is of order $5\%$. Note that we are not including a radiation component in this work.

Taking into account the model dependency on three parameters, namely $x$, $y$ and $z$, it is complicated to qualitatively describe the evolution of the relevant functions, therefore, in order to show the possible behavior of the density parameter associated to the torsional fluid we plot in figure \ref{fig5} some possible evolution of $\Omega_T$ for given values of $x$, $y$ and $z$. In the left panel the curves represent the evolution of $\Omega_T$ for different values of the parameter $x$ for fixed $y$ and $z$. We observe that $\Omega_T$ becomes negative in the past for $x>1.7$, which is not forbidden given that this density parameter corresponds to an effective fluid. For $1.5<x<1.7$ the torsional fluid contribution can remain important in the past and the matter domination period would experience important differences compared to the standard cosmological scenario. For $x<1.5$, the torsional fluid becomes the dominant contribution during a period that should correspond to matter domination. We notice that in general, the behaviors observed in this figure are common for different values of $y$ and $z$ in the allowed ranges. The right panel of figure \ref{fig5} is a closer view of $x=1.7$ and $y=0.5$ for different values of $z$, we observe that it is possible to have a small enough contribution of $\Omega_T$ in the past for some values of $z$, for instance, $z=0.55$.

As a particular example we have chosen a cosmological scenario given by the parameters $x=1.67$, $y=0.24$ and $z=0.56$. The evolution of the relevant cosmological functions of this scenario is shown in figures  \ref{figa} and \ref{figb}. In figure \ref{figa} we see the sign transition of the deceleration parameter $q$ at $\bar{a}\approx0.6$. The asymptotic limits of $q$ are $-1$ in the future and $0.5$ in the past, consistent with the standard cosmological scenario. On the other hand, we note that the Hubble expansion rate $H$ assumes the value $H_0$ twice because $\dot{H}$ becomes positive during the evolution with the sign-transition of $\dot{H}$ occurring in the future ($\bar{a}>1$).

Note that equation \eqref{scalefactor} imposes {$\bar{a}_0=1.33$} for {$\tau_0=0.54$} in order to have $\bar{a}(\tau_0)=1$ today. Besides, this scenario corresponds to an approximated evolution of 13.8 Gyr since $\bar{z}=200$, consistent with the standard cosmological evolution for $H_0=68$ [km/s/Mpc] \cite{Planck}.

In the left panel of figure \ref{figb} we show the behavior of the dimensionless density parameters, $\Omega_T=\frac{\rho_T}{3H^2}$ and $\Omega_m=\frac{\rho}{3H^2}$, where we have chosen the current value of the density parameter for pressureless matter to be {$\Omega_{m0}=0.3$}, according with the last measurements of the Planck satellite \cite{Planck}. We observe that around $\bar{a}\approx 0.3$ (or $\bar{z}\approx2$) the matter contribution corresponds to approximately $90\%$ of the overall, consistent with the standard cosmological scenario. The right panel of figure \ref{figb} shows the behavior of the state parameter for the torsional effective fluid $\omega_T=p_T/\rho_T$, this parameter is negative during the entire evolution and asymptotically tends to a cosmological constant in the future. It is interesting to note that among the many possible behaviors shown in figure  \ref{fig5}, we are able to find a combination of parameters where the standard cosmological scenario at late time seems to be reproduced at background level and the role of dark energy is played by an effective torsional fluid.

It is worth to pointing out that a scalar-tensor cosmological scenario as a single parameter extension of the standard cosmological model ($\Lambda$CDM) was presented in \cite{Algoner}. The authors found similar behaviors to those shown in the left panel of figure  \ref{fig5}, but representing the effective fluid for a non-torsional dark energy component. In this sense, it would be interesting to clarify if there is any distinctive imprint of the torsion on the scenarios here presented, which allows us to differentiate from the standard $\Lambda$CDM scenario and any other scalar-tensor cosmological scenario.

\subsection{Solution for $f=0$ with pressureless matter}
\label{SecT:3_2}

In this section we impose conditions (\ref{conditions}) along with $f=0$ into equations \eqref{Eq1}--\eqref{Eq5}. Specifically, by considering \eqref{Z0}, $N\neq0$ and assuming the matter content is a pressureless fluid, equations \eqref{Eq1}--\eqref{Eq3} become
\begin{eqnarray}
Z^2&=&-\frac{k}{a^2}+\frac{\beta}{3}+ \frac{\rho}{3 N},\label{EQ_4_34}\\
\dot Z&=& \frac{1}{2} \left(\beta-\frac{k}{a^2} -Z (2 H+Z)\right),\label{Z-eq}\\
\dot N &=& \frac{2 a^2 N (Z-H)}{a^2 \left(8 \alpha  Z^2+1\right)+8 \alpha  k}\label{dN}.
\end{eqnarray}
Then, by replacing \eqref{Z-eq} and \eqref{dN} into \eqref{Eq5}, for $8 \alpha \left(Z^2+\frac{k}{a^2}\right)+1\neq 0$, we obtain
\begin{equation}
\label{Eq_4_30}
(H-Z) \left(24 \alpha \left[Z^2+\frac{k}{a^2}\right]^2-3 \left[Z^2+\frac{k}{a^2}\right] (8 \alpha  \beta +1)-\beta\right) =0,
\end{equation}
which leads to the following two possibilities.
\begin{itemize}
\item $H=Z$, which implies $h=0$ and $N$ constant from \eqref{Z0} and \eqref{dN}, respectively. No torsional fluid is present in this case and given that the non-minimal coupling to the gravitational sector disappears in $\eqref{lagrangian}$, we recover the standard $\Lambda$CDM scenario, where the field equations, considering $N=1$, become
\begin{equation}
\label{LCDM}
\dot Z= \frac{1}{2} \left(-\frac{k}{a^2}+\beta -3 Z^2\right),\quad \dot a= a Z, \quad \dot\rho= -3 Z \rho , \quad Z^2+\frac{k}{a^2}=\frac{\beta}{3}+ \frac{\rho}{3}.
\end{equation}

\item For $H\neq Z$ (or $\dot{N}\neq0$), equation \eqref{Eq_4_30} implies
\begin{eqnarray}
Z^2+\frac{k}{a^2}+\frac{\beta}{3}-8\alpha \left(Z^2+\frac{k}{a^2}\right)\left(Z^2+\frac{k}{a^2}-\beta\right)=0,
\label{f6}
\end{eqnarray}
corresponding to a polynomial equation for $Z^2+\frac{k}{a^2}=Z_0^2$, where $Z_0$  is constant. By using this last result into (\ref{Eq2}) we get the following expansion rate,
\begin{eqnarray}
\label{Hf}
H=\pm\left(\frac{\beta-Z_0^2}{2Z_0^2}\right)\sqrt{Z_0^2-\frac{k}{a^2}},
\end{eqnarray}
valid for $Z_0>0$ and $\beta-Z_0^2>0$. By integrating equation (\ref{Hf}) for $Z>0$ we obtain the following scale factor,
\begin{eqnarray}
\label{scalefactorf0}
a(\tilde{\tau})=e^{\tilde{\tau}}+\frac{k e^{-\tilde{\tau}}}{4 Z_{0}^2}, \quad \textrm{where}\ \tilde{\tau}=\frac{\beta-Z_0^2}{2Z_0}(t-t_0),
\end{eqnarray}
where $t_0$ is an integration constant.
Furthermore, from equations \eqref{T3} and \eqref{EQ_4_34} we respectively obtain 
\begin{equation}
\rho (\tilde{\tau})= \rho_0 \left(e^{\tilde{\tau}}+\frac{k e^{-\tilde{\tau}}}{4 Z_{0}^2}\right)^{-3},\quad N(\tilde{\tau})= \left(\frac{\rho_0}{3Z_0^2-\beta}\right) \left(e^{\tilde{\tau}}+\frac{k e^{-\tilde{\tau}}}{4 Z_{0}^2}\right)^{-3},
\end{equation}
where $\rho_0$ is an integration constant.
\end{itemize}

We notice that the scale factor (\ref{scalefactorf0}) corresponds to eternal accelerated expansion, that is, this scenario does not allow a transition from decelerated to accelerated expansion even when pressureless matter is considered.
\section{Dynamical system analysis}\label{Sect:4}

In the section \ref{dos} we find an exact solution to equations \eqref{Eq1}--\eqref{Eq5}, for some specific model parameters and initial conditions. We noticed that the solution can correspond to a period of matter domination followed by accelerated expansion. Now, we are interested in studying the cosmological behavior in a general sense, independently of the initial conditions and the specific universe evolution. For this purpose we apply the dynamical systems approach, which allows to extract global features of the scenario at hand.
In this procedure, the first step is to transform the cosmological equations into an autonomous system, then find the fixed points and finally study their stability. For this end, linear perturbations around the critical points are performed, and the system is rewritten in terms of a perturbation matrix from which the type and stability of each critical point can be obtained. In the case of non-hyperbolic critical points, we should use the center manifold theorem or rely on numerical inspection \cite{Joint0,Joint1}.

\subsection{Dynamical system analysis for $f\neq0$ with presureless matter} \label{Sect:4.1}

We perform a dynamical systems analysis of the scenario defined by equations  (\ref{Eq1})--(\ref{Eq5}) considering $f\neq0$ and pressureless matter.
By replacing $W$ from equation \eqref{Eq1}, using the chain rule $\phi'(t) G'(\phi (t))=\dot G$ (for any function $G$ of $\phi(t)$) and imposing conditions \eqref{conditions}, from equations \eqref{Eq2}, \eqref{T3}, \eqref{Eq4} and \eqref{Eq3} we obtain respectively,
\begin{eqnarray}
\dot Z&=& \frac{1}{16} \left(\frac{1}{\alpha }+8 \beta -16 Z^2\right),\\
\dot\rho&=& \frac{\rho\left(N \left(8 \alpha  \beta -48 \alpha  Z^2+3\right)+8 \alpha  \rho\right)}{16 \alpha  Z N},\\
\dot N&=&  \frac{N}{8 \alpha  Z},\\
0&=&\frac{3}{8\alpha }+\beta +6 Z (H-Z)+\frac{\rho}{N},
\end{eqnarray}
where the last equation constitutes a constraint.

From equation \eqref{Z} we can easily note that, by defining a new variable $\zeta\equiv\frac{Z}{Z_e}$, it takes values over the interval $(-1,1)$ and it tends to $-1$ as $(t-t_i)\rightarrow -\infty$ and to $+1$ as $(t-t_i)\rightarrow \infty$. Note that here we are considering $Z_{e}=\sqrt{\frac{\beta}{2}+\frac{1}{16\alpha}}>0$. 

By defining the following dimensionless variables,
\begin{equation}
\label{vars_zeta}
U_1=-\frac{k}{a^2 Z_e^2 (1+ \zeta^2)},\quad U_2=\frac{\rho}{3N Z_e^2 (1+ \zeta^2)},\quad U_3=\frac{\zeta}{\sqrt{1+\zeta^2}},
\end{equation}
and a new time variable $\bar{\tau}$, in such a way that for any generic function $G$ we have
\begin{equation}
\label{newtime}
G'= \frac{d G}{d \bar{\tau}}=\frac{\zeta^2}{\left(\zeta^2+1\right)^{3/2}  Z_e} \dot G,
\end{equation}
we obtain the following autonomous system 
\begin{eqnarray}
U_1'&=&U_1 U_3 \left(-2 x +U_2+4 U_3^4+2 x  U_3^2-6 U_3^2+2\right),\label{eq3.37}\\
U_2'&=&\frac{1}{2} U_2 U_3 \left(3 U_2+8 U_3^4-12 U_3^2+2\right),  \label{eq3.38}\\
U_3'&=&U_3^2(1-U_3^2)\left(1-2 U_3^2\right), 
\label{eq3.39}
\end{eqnarray}
which defines a flow on the phase space defined by
\begin{equation}
\label{ps}
\left\{(U_1,U_2,U_3)\in\mathbb{R}^3:U_1+U_2\leq (x +1) U_3^2 -x,\quad-\frac{\sqrt{2}}{2}\leq U_3 \leq \frac{\sqrt{2}}{2}\right\},
\end{equation}
where the first inequality comes from the reality condition $\frac{f^2}{Z^2}\geq 0$ imposed into equation (\ref{Eq1}) and the second from the definition of $\zeta$. Note that, we use the dimensionless parameter $x=\beta Z_{e}^{-2}/3$ (see table \ref{table1} for a description of its role in the dynamics). 

\subsubsection{Fixed points at the finite region}
\label{Sect:4.1.1}
In this section we find the fixed points and study their stability by examining the eigenvalues of the perturbation matrix, the main results are summarized in table \ref{table2}. The fixed points corresponding to $U_3^2=1$ are not included in table \ref{table2} because they lie outside of the physical portion of the phase space \eqref{ps}.
\begin{table}[ht!]
\centering
\scalebox{0.9}{
\begin{tabular}{|c|c|c|c|c|}
\hline
Labels & $(U_1,U_2,U_3)$ & 
Existence & Stability & $(q,\Omega_{m}^{\text{eff}},\Omega_{k}^{\text{eff}},\omega_{total})$ \\\hline
$P_1$ & $\left(0,\frac{2}{3},-\frac{1}{\sqrt{2}}\right)$ &  $x \leq -\frac{1}{3}$& saddle& $\left(-1,\frac{12 }{(2-3 x )^2},0,-1\right)$ \\ \hline
$P_2$& $\left(0,\frac{2}{3}, \frac{1}{\sqrt{2}}\right)$ & $x \leq -\frac{1}{3}$& saddle& $\left(-1,\frac{12 }{(2-3 x )^2},0,-1\right)$ \\\hline
$P_3$ & $\left(0,0,-\frac{1}{\sqrt{2}}\right)$ & $x<1$ & $\begin{array}{c} \text{saddle for}\;x<0 \\ \text{source for}\; 0<x< 1
\end{array}$& $(-1,0,0,-1)$ \\ \hline
$P_4$ & $\left(0,0, \frac{1}{\sqrt{2}}\right)$ & $x<1$ &  $\begin{array}{c} \text{saddle for}\; x<0\\ \text{sink for}\; 0<x< 1\end{array}$& $(-1,0,0,-1)$ \\ \hline 
$P_5$ & $\left(U_{1c}, U_{2c}, 0\right)$ & $U_{1c}+U_{2c}\leq -x$& non-hyperbolic &
$\Big(\frac{2 \left(-2 (x -2) (x -1)+U_{2c}^2+2 (x -1) U_{2c}\right)}{(-2 x +U_{2c}+2)^2}, 0,0, $\\
&&&&$ \frac{-4 (x -1) (3 x -5)+3 U_{2c}^2+12 (x -1) U_{2c}}{3 (-2 x +U_{2c}+2)^2}\Big)$ \\\hline
\end{tabular}}
\caption{\label{table2} Fixed points, existence and stability conditions for the dynamical system defined by equations \eqref{eq3.37}--\eqref{eq3.39}, restricted to the phase space \eqref{ps}.}
\end{table}
To characterize the solutions given by the fixed points in table \ref{table2} we define the following variables:
\begin{eqnarray*}
\Omega_m^{\text{eff}}=\frac{\rho}{3 N H^2},\quad
\Omega_{k}^{\text{eff}}= -\frac{k }{a^2 H},\quad
\omega_{total}=\frac{p_{total}}{\rho_{total}}.
\end{eqnarray*}

For $0< x<1$ the early and late-time attractors correspond to a de Sitter solution, representing an accelerated stage. For $x<0$, the solutions are all saddles such that the acceleration phase is a transient phenomena. Since almost all the fixed points have $q=-1$, the acceleration (at early, late or intermediate epochs) is a generic feature of our model. 

On the other hand, we have not seen any fixed point at the finite region of the phase space corresponding to matter domination. However, for example, for the choice $U_{1c}=0, U_{2c}=2-2 x - 2 \sqrt{\frac{2}{3}} \sqrt{(x -1) (3 x -4)}+2$, the set $P_5$ in table \ref{table2} contains one point with $(q,  \Omega_{m}^{\text{eff}},\Omega_{k}^{\text{eff}},\omega_{total})=\left(\frac{1}{2},0,0,0\right)$, such that the torsional fluid can mimic dust for $x<\frac{2}{3}$ or $x>\frac{10}{7}$. Since in this case all the eigenvalues are zero, we rely on numerical experiments to show the stability. 

\subsubsection{Fixed points at infinity}\label{fixpinf}

Observe that the dynamical system given by equations \eqref{eq3.37}--\eqref{eq3.39} admits several fixed points in the finite region, whose stability can be obtained by following the usual linearization procedure and examining the eigenvalues of the involved perturbation matrix. However, since the dynamical system is defined on unbounded phase spaces, there may exist non-trivial behaviors at the region where the variables diverge. For this reason, we need to introduce Poincar\'e variables to study the behavior of the system at infinity. 

Since the variable $U_3$ is bounded, the compactification of the phase space can be achieved by setting:
\begin{eqnarray}
U_1=\bar{r} \cos\psi,\quad U_2=\bar{r} \sin \psi,\quad U_3= \frac{R}{\sqrt{2}},
\end{eqnarray}
and $\bar{r}=\frac{r}{1-r}$, where $0\leq \bar r< \infty$, $-1\leq R\leq 1$, and $0\leq \psi \leq 2\pi$. It is interesting to note that in the limit $r\rightarrow 1^{-}$ ($\bar{r}\rightarrow \infty$) the leading terms of the set of equations \eqref{eq3.37}--\eqref{eq3.39} take the following form
\begin{align}
& r'\rightarrow \frac{R (11 \sin (\psi )-\sin (3 \psi ))}{8 \sqrt{2}}, \label{eq3.55}\\
&\psi' \rightarrow \frac{R \sin ^2(\psi ) \cos (\psi )}{2 \sqrt{2}}, \label{eq3.56}\\
&R'\rightarrow\frac{R^2 \left(R^4-3 R^2+2\right)}{2 \sqrt{2}}. \label{eq3.57}
\end{align}
We note that we can find the equilibrium points at infinity just by using equations \eqref{eq3.56}--\eqref{eq3.57}, considering $R\neq0$ and setting $\psi'=0$ and $R'=0$. 

\begin{table}[ht!]
\centering
\scalebox{1}{
\begin{tabular}{|c|c|c|c|}
\hline
Labels & $(r, R, \psi, {X}_1, {X}_2, U_3)$ & $\{\lambda_1, \lambda_2, \lambda_3\}$ & Stability  \\\hline
$Q_1$ & $\left(1, -1, 0, 1,0,-\frac{1}{\sqrt{2}}\right)$ & $\left\{\frac{1}{\sqrt{2}},0,0\right\}$ & non-hyperbolic  \\\hline
$Q_2$ & $\left(1,-1,\frac{\pi}{2},0,1,-\frac{1}{\sqrt{2}}\right)$ & $\left\{\frac{1}{\sqrt{2}},\frac{1}{2 \sqrt{2}},-\frac{3}{2 \sqrt{2}}\right\}$ & source  \\\hline
$Q_3$ & $\left(1,-1,\pi,-1,0,-\frac{1}{\sqrt{2}}\right)$ & $\left\{\frac{1}{\sqrt{2}},0,0\right\}$ & non-hyperbolic  \\\hline
$Q_4$ & $\left(1,-1,\frac{3\pi}{2},0,-1,-\frac{1}{\sqrt{2}}\right)$ & $\left\{\frac{1}{\sqrt{2}},-\frac{1}{2 \sqrt{2}},\frac{3}{2 \sqrt{2}}\right\}$ & saddle \\\hline 
$Q_5$ & $\left(1, 1, 0, 1, 0,\frac{1}{\sqrt{2}}\right)$ &$\left\{-\frac{1}{\sqrt{2}},0,0\right\}$ & non-hyperbolic \\\hline 
$Q_6$ & $\left(1, 1, \frac{\pi}{2}, 0,1,\frac{1}{\sqrt{2}}\right)$ & $\left\{-\frac{1}{\sqrt{2}},-\frac{1}{2 \sqrt{2}},\frac{3}{2 \sqrt{2}}\right\}$ & sink \\\hline
$Q_7$ & $\left(1, 1, \pi, -1,0,\frac{1}{\sqrt{2}}\right)$ &$\left\{-\frac{1}{\sqrt{2}},0,0\right\}$ & non-hyperbolic \\\hline
$Q_8$ & $\left(1, 1, \frac{3\pi}{2}, 0,-1,\frac{1}{\sqrt{2}}\right)$ & $\left\{-\frac{1}{\sqrt{2}},\frac{1}{2 \sqrt{2}},-\frac{3}{2 \sqrt{2}}\right\}$ &  saddle  \\\hline 
\end{tabular}}
\caption{\label{table3} Fixed points, eigenvalues and stability for the dynamical system defined by equations \eqref{eq3.55}--\eqref{eq3.57}.  We have defined $X_1=r\cos\psi, X_2=r\sin\psi$ for $R\neq0$.}
\end{table}

The fixed points corresponding to equations \eqref{eq3.55}--\eqref{eq3.57} are presented in table \ref{table3}. The fixed points at infinity having $R^2=2$ are not considered because they lie outside of the physical portion of the phase space.

A given fixed point at infinity is a sink if $\lambda_1=\frac{R \left(3 R^4-6 R^2+2\right)}{\sqrt{2}}<0, \lambda_2=\frac{R \sin (\psi ) (3 \cos (2 \psi )+1)}{4 \sqrt{2}}<0$, $\lambda_3=\frac{R (11 \sin (\psi )-\sin (3 \psi ))}{8 \sqrt{2}}>0$, where $\lambda_1$ and $\lambda_2$ are the eigenvalues of the matrix 
\begin{equation}
\left(
\begin{array}{cc}
 \frac{\partial \psi'}{\partial \psi} & \frac{\partial \psi'}{\partial R} \\      
  \frac{\partial R'}{\partial \psi} & \frac{\partial R'}{\partial R}\\
\end{array}
\right)=\left(
\begin{array}{cc}
 \frac{R \sin (\psi )(1+3\cos(2\psi))}{4\sqrt{2}}& \frac{\sin ^2(\psi )\cos (\psi ) }{2 \sqrt{2}} \\
 0 & \frac{R(2-6R^2+3R^4)}{\sqrt{2}} \\
\end{array}
\right),
\end{equation}
meanwhile the condition on $\lambda_3$ corresponds to $r'>0$ which means that $r$ is a monotonic increasing function that tends to $1$ from below. On the other hand, a given fixed point corresponds to a source if $\lambda_1>0, \lambda_2>0, \lambda_3<0$ and it is a saddle otherwise (for non-zero $\lambda_1$, $\lambda_2$ and $\lambda_3$).

In the case $R=0$, the leading terms of \eqref{eq3.55}--\eqref{eq3.56} as $r\rightarrow 1^{-}$ and $R\rightarrow0^+$ take the following form
\begin{align}
& r'\rightarrow \frac{ (11 \sin (\psi )-\sin (3 \psi ))}{8 }, \label{eq4-32}\\
&\psi' \rightarrow \frac{\sin ^2(\psi ) \cos (\psi )}{2 }. \label{eq4-33}
\end{align}
We note that the equilibrium points at infinity are found just by using equation \eqref{eq4-33} and setting $\psi'=0$. 
To analyze the stability of the points at infinity we define
$\mu_1:= \frac{\partial \psi'}{\partial \psi}= \frac{3 \sin (3 \psi )-\sin (\psi )}{8}$ and $\mu_2= \frac{ 11 \sin (\psi )-\sin (3 \psi )}{8 } $, then a given fixed point at infinity is a sink if $\mu_1<0, \mu_2>0$, it corresponds to a source if $\mu_1>0,\ \mu_2<0$ and it is a saddle otherwise (for non-zero $\mu_1$, and $\mu_2$). The corresponding fixed points at infinity for the case $R=0$ are presented in table \ref{tableInfinityU3zero}.

\begin{table}[h!]
\centering
\scalebox{1}{
\begin{tabular}{|c|c|c|c|}
\hline
Labels & $(r, \psi, {X}_1, {X}_2)$ & $\{\mu_1, \mu_2\}$ & Stability  \\\hline
$W$ & $\left(1, 0, 1, 0\right)$ &$\left\{0,0\right\}$ & non-hyperbolic\\\hline 
$N$ & $\left(1, \frac{\pi}{2}, 0,1\right)$ & $\left\{-\frac{1}{2},\frac{3}{2}\right\}$ & sink \\\hline
$E$ & $\left(1, \pi, -1,0\right)$ &$\left\{0,0\right\}$ & non-hyperbolic \\\hline
$S$ & $\left(1, \frac{3\pi}{2}, 0,-1\right)$ & $\left\{\frac{1}{2},-\frac{3}{2}\right\}$ &  source  \\\hline 
\end{tabular}}
\caption{\label{tableInfinityU3zero} Fixed points, eigenvalues and stability for the dynamical system defined by equations \eqref{eq4-32}--\eqref{eq4-33}. We have  defined $X_1=r\cos\psi, X_2=r\sin\psi$ for $R=0$. }
\end{table}
\subsubsection{Invariant sets}\label{Sect:4.1.2}

In this section we present several invariant sets in which the analysis is simpler and we discuss some numerical integrations.

The set $U_3=0$ defines an invariant set that cannot be crossed by any solution, such that the phase space is divided in two sectors: the sector $\Gamma_{-}$ with $U_3\in\left[-\frac{\sqrt{2}}{2}, 0\right)$ and the sector $\Gamma_+$ with $U_3\in\left(0,\frac{\sqrt{2}}{2}\right]$.  From equation \eqref{eq3.39} follows that $U_3$ is a monotonic function taking values on the interval $\left[-\frac{\sqrt{2}}{2}, 0\right) \cup \left(0,\frac{\sqrt{2}}{2}\right]$, and  hence $U_3$ can be viewed
as a time variable if one is so inclined. 

We notice that by integrating equation \eqref{eq3.39} we find for $U_3\neq 0$ 
\begin{equation}
\bar{\tau}-\bar{\tau}_0= -\frac{1}{U_3}-\tanh ^{-1}(U_3)+2 \sqrt{2} \tanh ^{-1}\left(\sqrt{2} U_3\right).
\end{equation}
Hence $\bar{\tau}\rightarrow -\infty$ as $U_3\rightarrow -\frac{\sqrt{2}}{2}$ and $\bar{\tau}\rightarrow +\infty$ as $U_3\rightarrow 0$ in the sector $\Gamma_-$ and, $\bar{\tau}\rightarrow -\infty$ as $U_3\rightarrow 0$ and $\bar{\tau}\rightarrow +\infty$ as $U_3\rightarrow \frac{\sqrt{2}}{2}$ in the sector $\Gamma_+$. 

On the other hand, the system of equations \eqref{eq3.37}--\eqref{eq3.39} is integrable on $\Gamma_- \cup \Gamma_+$. Indeed, for $U_3\neq 0$ we obtain 
\begin{align}
&\frac{d U_1}{d U_3}=\frac{U_1 \left(-2 x +U_2+4 U_3^4+2 x  U_3^2-6 U_3^2+2\right)}{U_3(1-U_3^2) \left(1-2 U_3^2\right)},\label{Neweq3.37}\\
&\frac{d U_2}{d U_3}=\frac{ U_2  \left(3 U_2+8 U_3^4-12 U_3^2+2\right)}{2U_3(1-U_3^2) \left(1-2 U_3^2\right)},  \label{Neweq3.38}
\end{align}
then by integrating and using the last relation in \eqref{vars_zeta}, we get the general solution 
\begin{align}
\label{sol1}
&U_1= \frac{c_2 \zeta ^{2(1- x)} \left(1-\zeta
   ^2\right)^x}{\left(1+\zeta ^2\right) \left(2 c_1-3 \zeta\right){}^{2/3}}, \quad 
&U_2= \frac{2\zeta (1-\zeta^2) }{\left(1+\zeta ^2\right) \left(2 c_1-3 \zeta\right)},  \quad \zeta\in[-1,0) \cup (0,1],
\end{align}
where $c_1$ and $c_2$ are integration constants. 
Using the definitions of $U_1$ and $U_2$ given by \eqref{vars_zeta} we finally obtain 
\begin{align}
\label{sol2}
&-\frac{k}{a^2 Z_e^2}= \frac{c_2 \zeta ^{2(1- x)} \left(1-\zeta
   ^2\right)^x}{\left(2 c_1-3 \zeta\right){}^{2/3}},\quad 
&\frac{\rho}{3 N Z_e^2}= \frac{2\zeta (1-\zeta^2) }{\left(2 c_1-3 \zeta\right)},  \quad \zeta\in[-1,0) \cup (0,1].
\end{align}

In the following we study the asymptotic behavior  of the sectors $\Gamma_+$ and $\Gamma_-$.
\paragraph{Dynamics in the sector $\Gamma_+$.}\label{Sect:4.1.3}

In this sector of the phase space $U_3>0$, so that the dynamics is restricted to
\begin{equation}
\left\{(U_1,U_2,U_3)\in\mathbb{R}^3:U_1+U_2\leq (x +1) U_3^2 -x, \quad 0\leq U_3 \leq \frac{\sqrt{2}}{2}\right\}.
\end{equation}
The late-time attractor in this sector is located at the invariant set $U_3=\frac{\sqrt{2}}{2}$ and then, the dynamics in this invariant set is governed by  equations \eqref{eq3.37} and \eqref{eq3.38} in the form:
\begin{align}
&U_1'=\frac{U_1 (U_2-x)}{\sqrt{2}},\label{eq4.32}\\
&U_2'=\frac{U_2 (3 U_2-2)}{2 \sqrt{2}},\label{eq4.33}
\end{align}
defined in the phase space
\begin{equation}
\left\{
(U_1,U_2)\in \mathbb{R}^2: x\leq 1 -2 U_1-2 U_2
\right\}.
\end{equation}

\begin{figure}[ht!]
\centering
\includegraphics[width=1.00\textwidth]{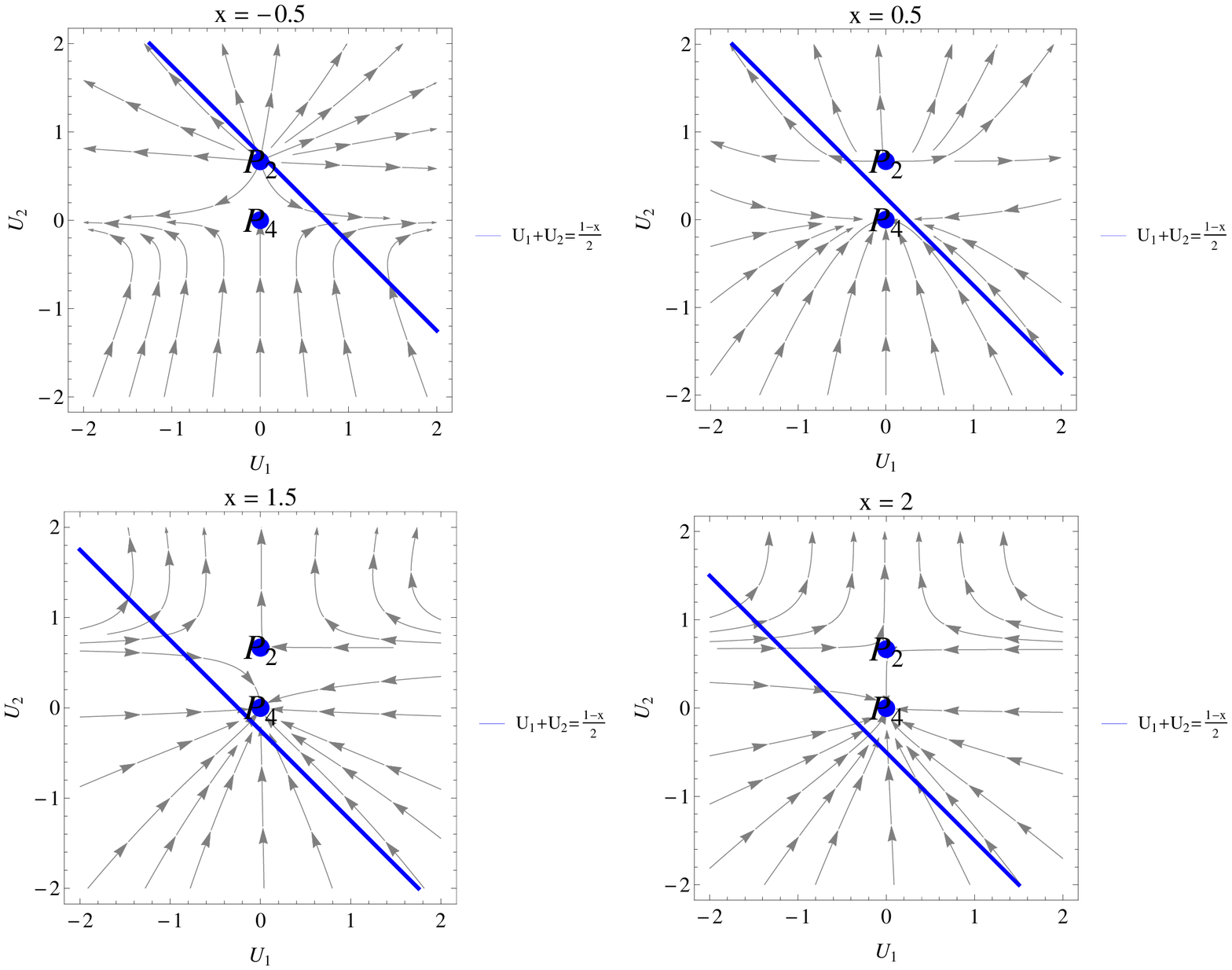}
\caption{\label{fig:DS3} Stream plot of the vector field defined by equations \eqref{eq4.32} and \eqref{eq4.33}, which gives the dynamics at the invariant set $U_3=\frac{\sqrt{2}}{2}$ for $x<0$, $0<x<1$, $1<x<2$ and $x=2$. The region below the blue line $ x= 1 -2 U_1-2 U_2$ denotes the physical region of the phase space.}
\end{figure}

\begin{figure}[ht!]
\centering
\includegraphics[width=1.00\textwidth]{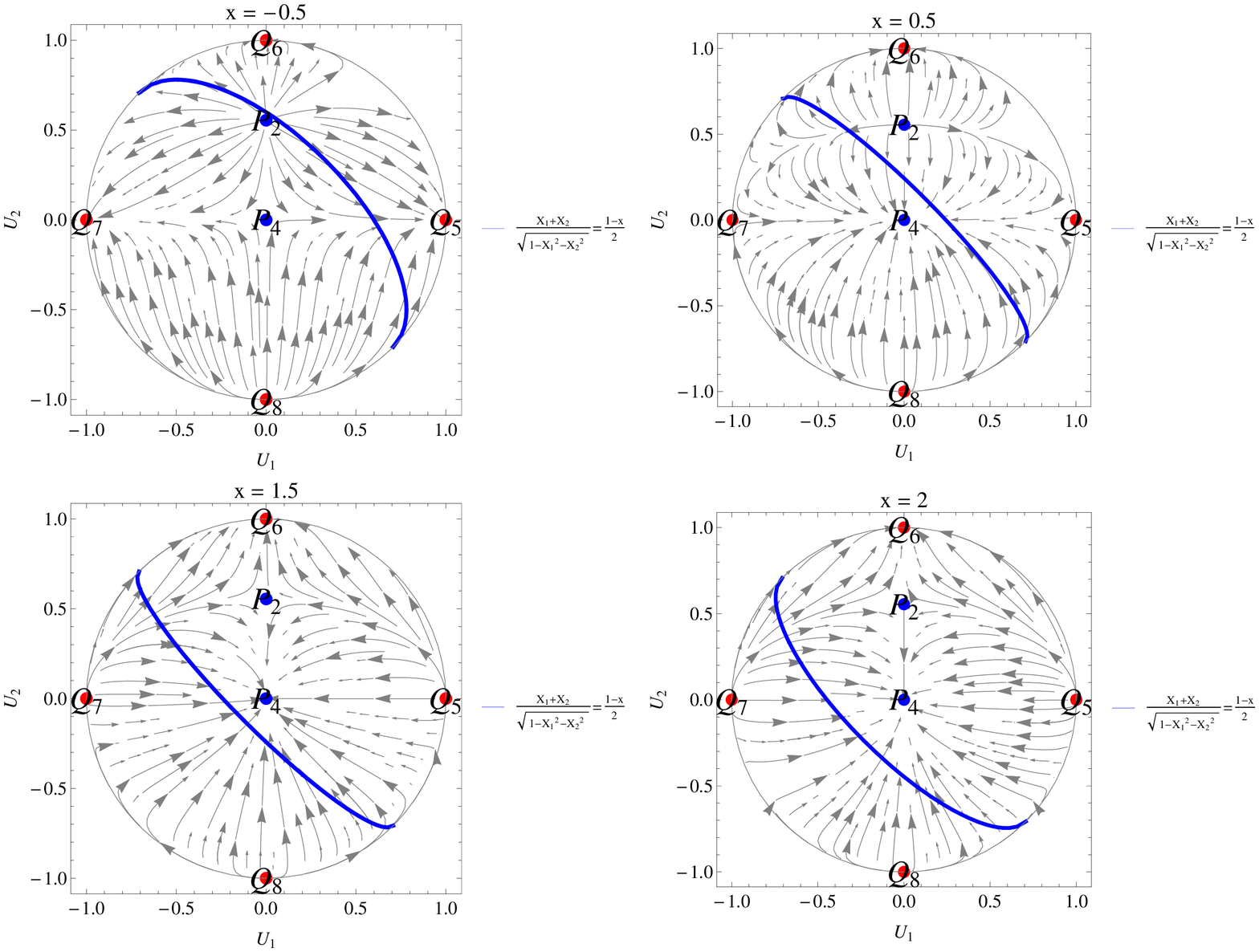}
\caption{\label{fig:DS4} Representation in Poincar\'e variables (see section \ref{fixpinf}) of the vector field defined by equations \eqref{eq4.32} and \eqref{eq4.33}, which gives the dynamics at the invariant set $U_3=\frac{\sqrt{2}}{2}$ for  $x<0$, $0<x<1$, $1<x<2$ and $x=2$. The region below the blue line denotes the physical region of the phase space. }
\end{figure}

In figure \ref{fig:DS3} we present stream plots of the vector field defined by equations \eqref{eq4.32} and \eqref{eq4.33} for the following choices: $x<0$, $0<x<1$, $1<x<2$ and $x=2$. The fixed points shown, $P_2$ and $P_4$, are defined in table \ref{table2} and the region below the blue line $ x= 1 -2 U_1-2 U_2$ denotes the physical region of the phase space (hereafter, we refer by ``a fixed point belongs to the physical region'' to the fact that the reality condition $\frac{f^2}{Z^2}\geq 0$ is satisfied at the fixed point). In this figure we see that, for $0<x<1$ the late-time attractor is $P_4$ corresponding to a de Sitter solution, meanwhile for $x \geq 1$ the attractor is $P_4$ but it lies outside the physical region of the phase space. Note that $P_2$ is generically a saddle for the three-dimensional dynamical system \eqref{eq3.37}--\eqref{eq3.39} (see table \ref{table2}), however, when the dynamics is restricted to the invariant set $U_3=\frac{\sqrt{2}}{2}$, it can be either a source for $x<\frac{2}{3}$ or a saddle for $x>\frac{2}{3}$. In figure \ref{fig:DS3} we also observe an open set of orbits going to infinity.

In figure \ref{fig:DS4} it is presented the projection in Poincar\'e variables showing the global dynamics at both finite and infinite regions for the invariant set $U_3=\frac{\sqrt{2}}{2}$. The fixed points shown in this figure, $Q_5,\ Q_6,\ Q_7,\ Q_8$ are defined in table \ref{table3}. Note that $Q_8$ is generically a saddle at infinity for the three-dimensional dynamical system \eqref{eq3.55}--\eqref{eq3.57}, however, when the dynamics is restricted to the invariant set $U_3=\frac{\sqrt{2}}{2}$, it is always a source.

\begin{figure}[ht!]
\centering
\includegraphics[width=0.5\textwidth]{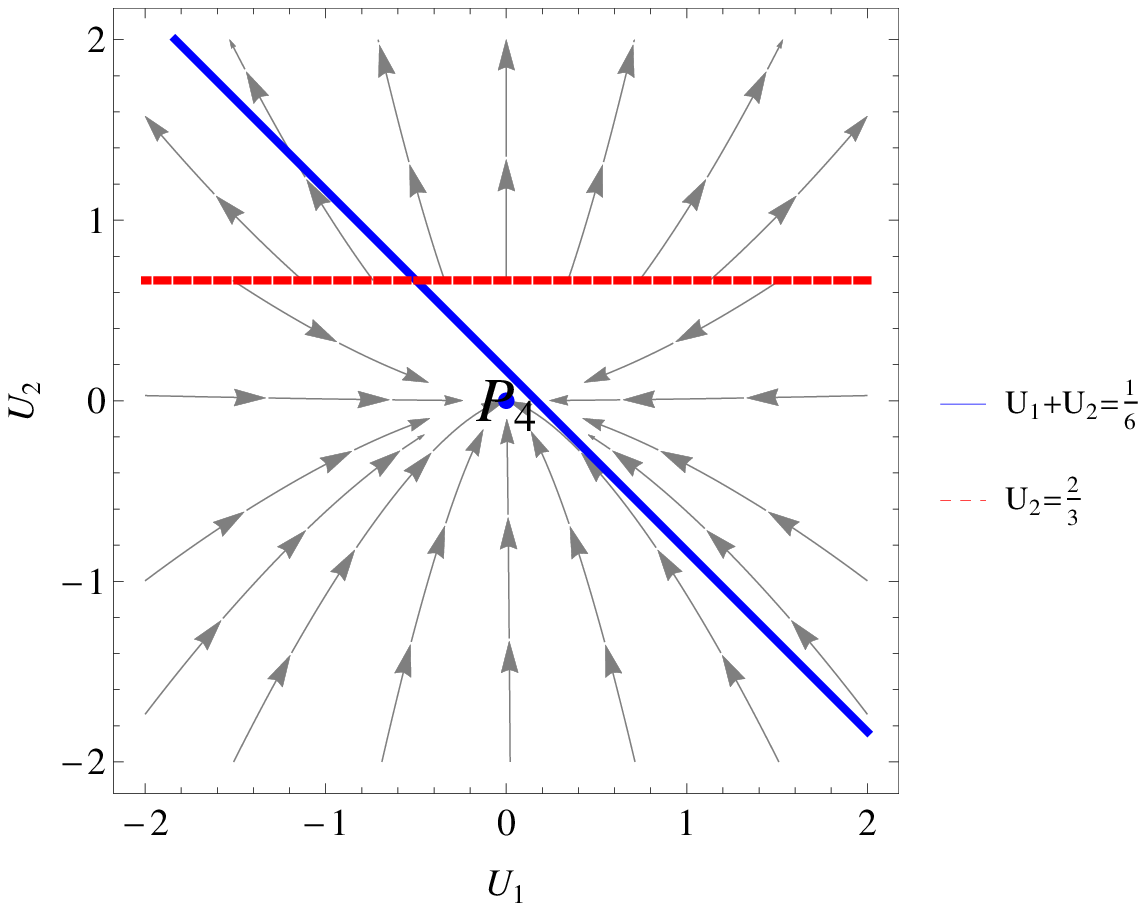}
\includegraphics[width=0.4\textwidth]{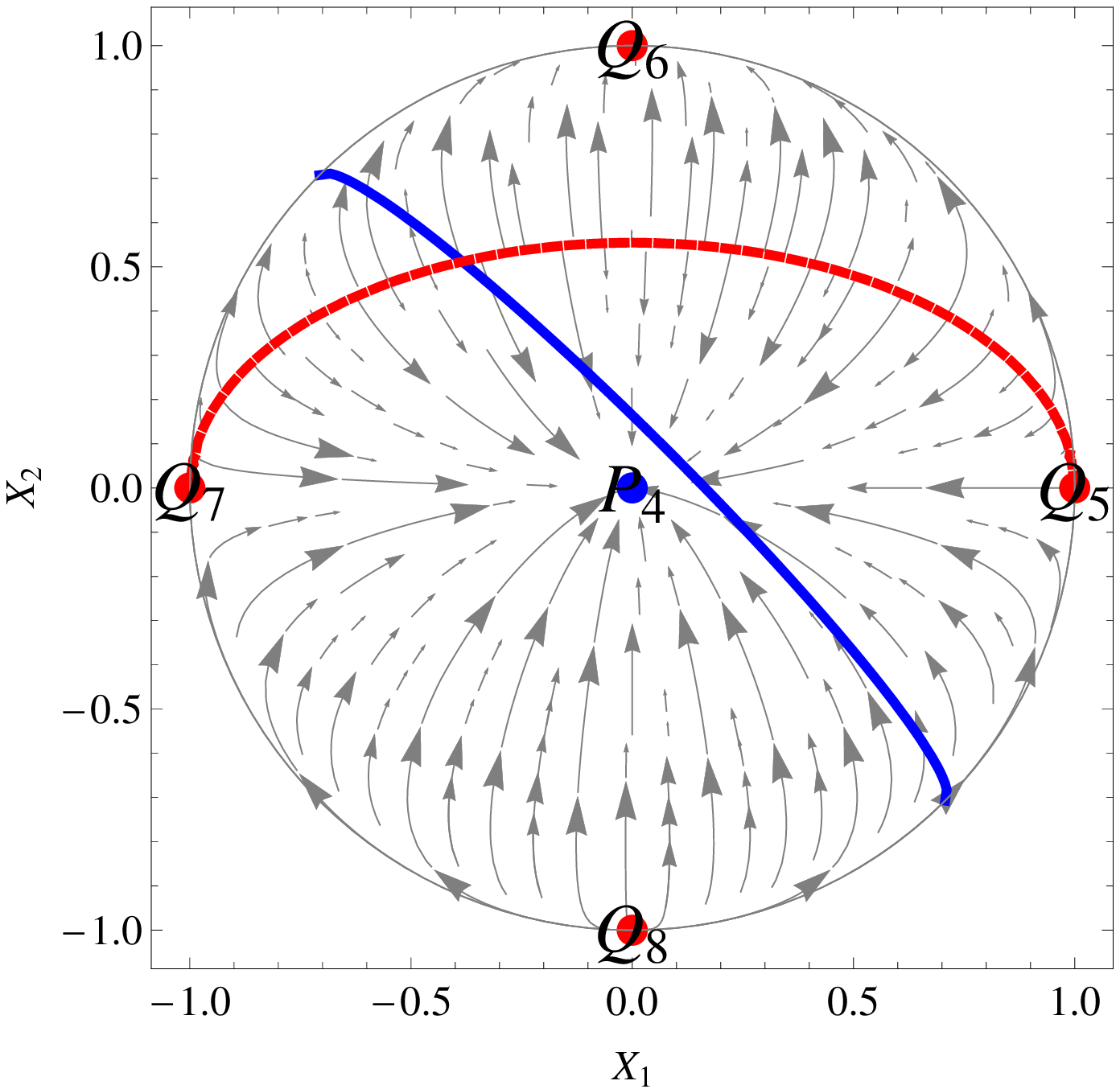}
\caption{\label{fig:DS1}Stream plot of the vector field defined by equations \eqref{eq4.32} and \eqref{eq4.33} for $x=\frac{2}{3}$,  showing the dynamics at the invariant set $U_3=\frac{\sqrt{2}}{2}$. The right panel shows the representation in the Poincar\'e variables (see section \ref{fixpinf}). The region below the blue line $U_1+U_2=\frac{1}{6}$ represents the physical portion of the phase space. The red dashed line represents the line of fixed points $P_2$.}
\end{figure}

Observe that the scenario presented in \cite{Zanelli-Toloza} is recovered from equations \eqref{eq3.37}--\eqref{eq3.39} by setting $x=\frac{2}{3}$. Nevertheless, in this case the fixed points $P_1$ and $P_2$ in table \ref{table2} are no longer single fixed points, but instead they become lines of fixed points, with $U_1=U_{1c},\ U_2=\frac{2}{3}$ and $U_3=\pm\frac{1}{\sqrt{2}}$. 
A stream plot on the invariant set $U_3=\frac{\sqrt{2}}{2}$ for the scenario \cite{Zanelli-Toloza} is presented in figure \ref{fig:DS1}, where the left panel shows the critical point at the finite region and the right panel shows the corresponding representation in Poincar\'e variables (see section \ref{fixpinf}), the red dashed line represents the critical point $P_2$ and the region below the thick blue line represents the physical portion of  the phase space. In the scenario \cite{Zanelli-Toloza} the future attractor in the finite region of the phase space is $P_4$, $Q_6$ is the attractor in the region at infinity, $Q_5$ and $Q_7$ are saddles and $Q_8$ is a source (remind $Q_8$ is a saddle for the three-dimentional dynamical system). In the physical portion of the phase space there are typical orbits that connect $P_2 \rightarrow Q_7 \rightarrow P_4$ (de Sitter solution) and typical orbits that connect $Q_8 \rightarrow Q_7 \rightarrow P_4$ (de Sitter solution).

On the other hand, the past  attractor in $\Gamma_+$ is located at the invariant set  $U_3=0$. The dynamics in this invariant set is governed by equations \eqref{eq3.37} and \eqref{eq3.38} in the form:
\begin{eqnarray}
U_1'&=&U_1 \left(-2 x +U_2+2\right),\label{eq4.39}\\
U_2'&=&\frac{1}{2} U_2  \left(3 U_2+2\right),\label{eq4.40}
\end{eqnarray}
defined in the unbounded phase space
\begin{equation}
\left\{
(U_1,U_2)\in \mathbb{R}^2:U_1+U_2\leq  -x
\right\}.
\end{equation}
The fixed point in this invariant set are given by:
\begin{itemize}
\item $P_{5a}: \left(U_1, U_2\right)=(0,0)$. In this case the eigenvalues are $\{1, 2 (1- x)\}$ and this point belongs to the physical region of the phase space for $x\leq 0$, such that this point corresponds to a source. 
\item $P_{5 b}: \left(U_1, U_2\right)=\left(0,-\frac{2}{3}\right)$. In this case the eigenvalues are $\left\{-1,-\frac{2}{3} (3 x-2)\right\}$ and this point belongs to the physical region of the phase space for  $x\leq \frac{2}{3}$, such that this point corresponds to a saddle.
\end{itemize}

\begin{figure}[ht!]
\centering
\includegraphics[width=1.00\textwidth]{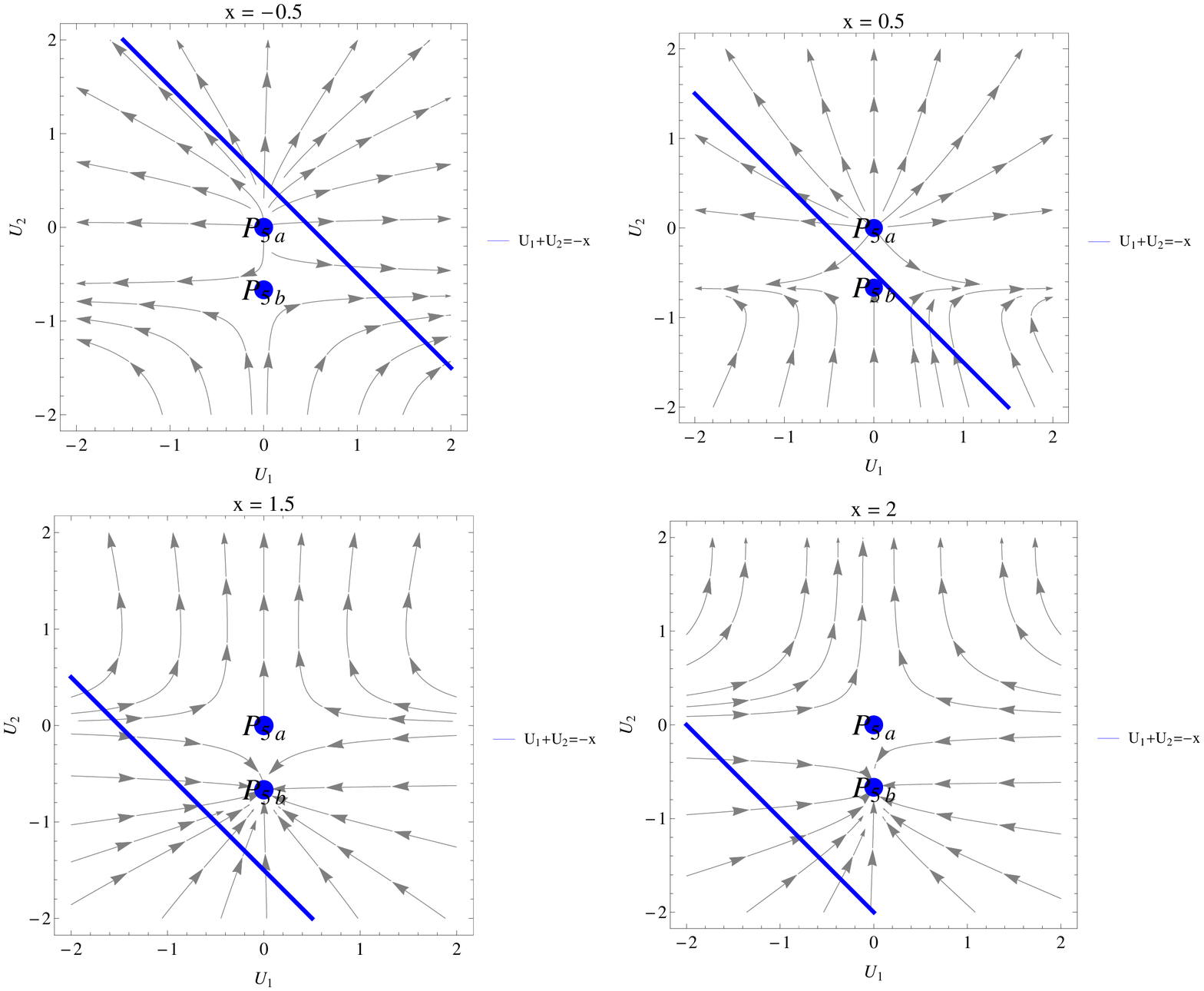}
\caption{\label{fig:DS7} Stream plot of the vector field defined by equations \eqref{eq4.39} and \eqref{eq4.40},  which gives the dynamics at the invariant set $U_3=0$ for $x<0$, $0<x<1$, $1<x<2$ and $x=2$. The region below the blue line $  U_1+ U_2=-x$ denotes the physical region of the phase space.}
\end{figure}

\begin{figure}[ht!]
\centering
\includegraphics[width=1.00\textwidth]{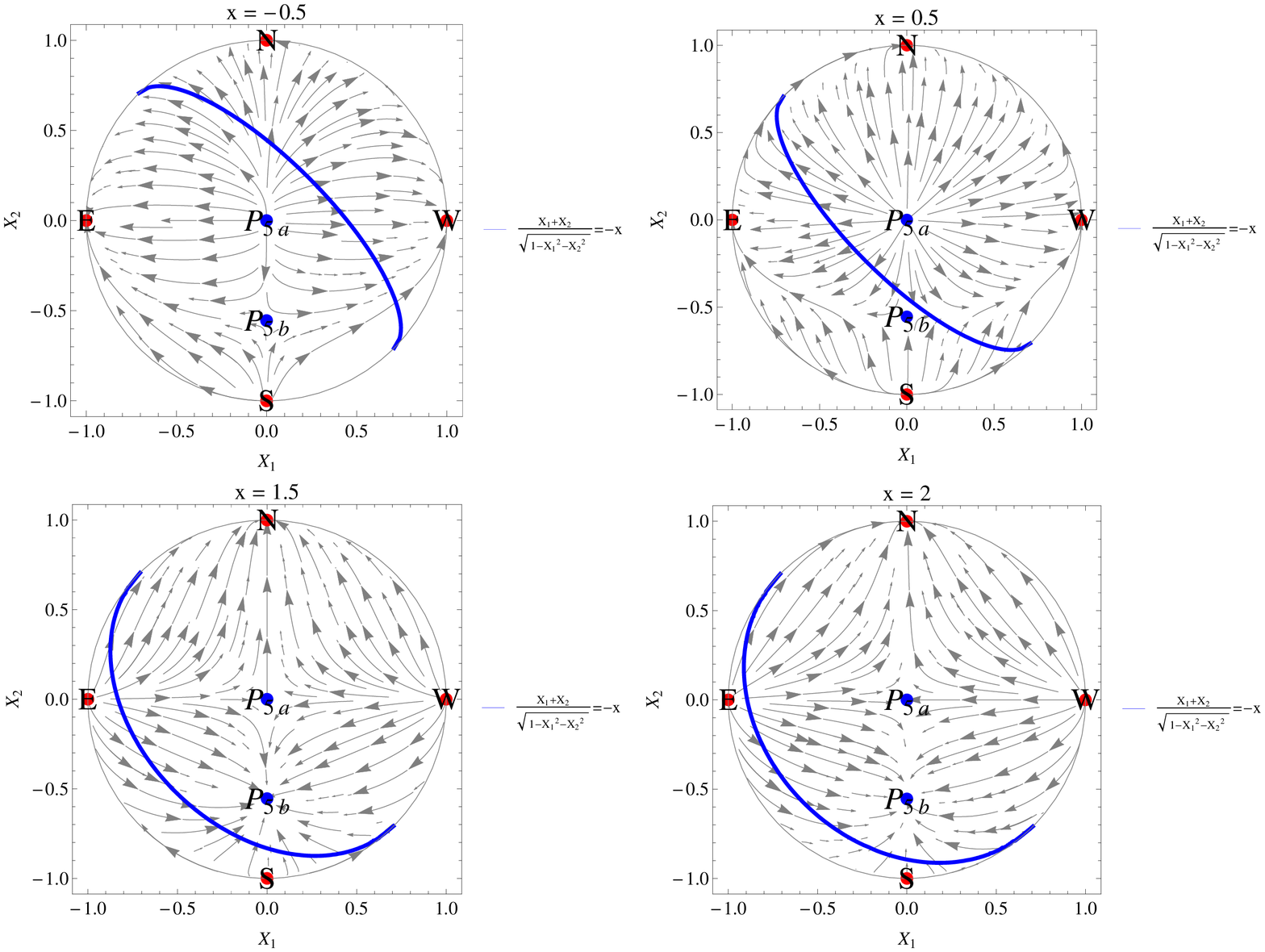}
\caption{\label{fig:DS8} Representation in Poincar\'e variables (see section \ref{fixpinf}) of the vector field defined by equations \eqref{eq4.39} and \eqref{eq4.40},  which gives the dynamics at the invariant set $U_3=0$ for $x<0$, $0<x<1$, $1<x<2$ and $x=2$. The region below the blue line denotes the physical region of the phase space.}
\end{figure}

In figure \ref{fig:DS7} it is shown a stream plot of the vector field defined by equations \eqref{eq4.39} and \eqref{eq4.40},  which gives the dynamics at the invariant set $U_3=0$ for the following choices: $x<0$, $0<x<1$, $1<x<2$ and $x=2$. The region below the blue line $  U_1+ U_2=-x$ denotes the physical region of the phase space. Figure \ref{fig:DS8} shows the corresponding representation in Poincar\'e variables for the invariant set $U_3=0$, the critical points at inifinity $N$, $S$, $E$ and $W$ are presented in table \ref{tableInfinityU3zero}. 

\begin{figure}[ht!]
\centering
\includegraphics[width=0.5\textwidth]{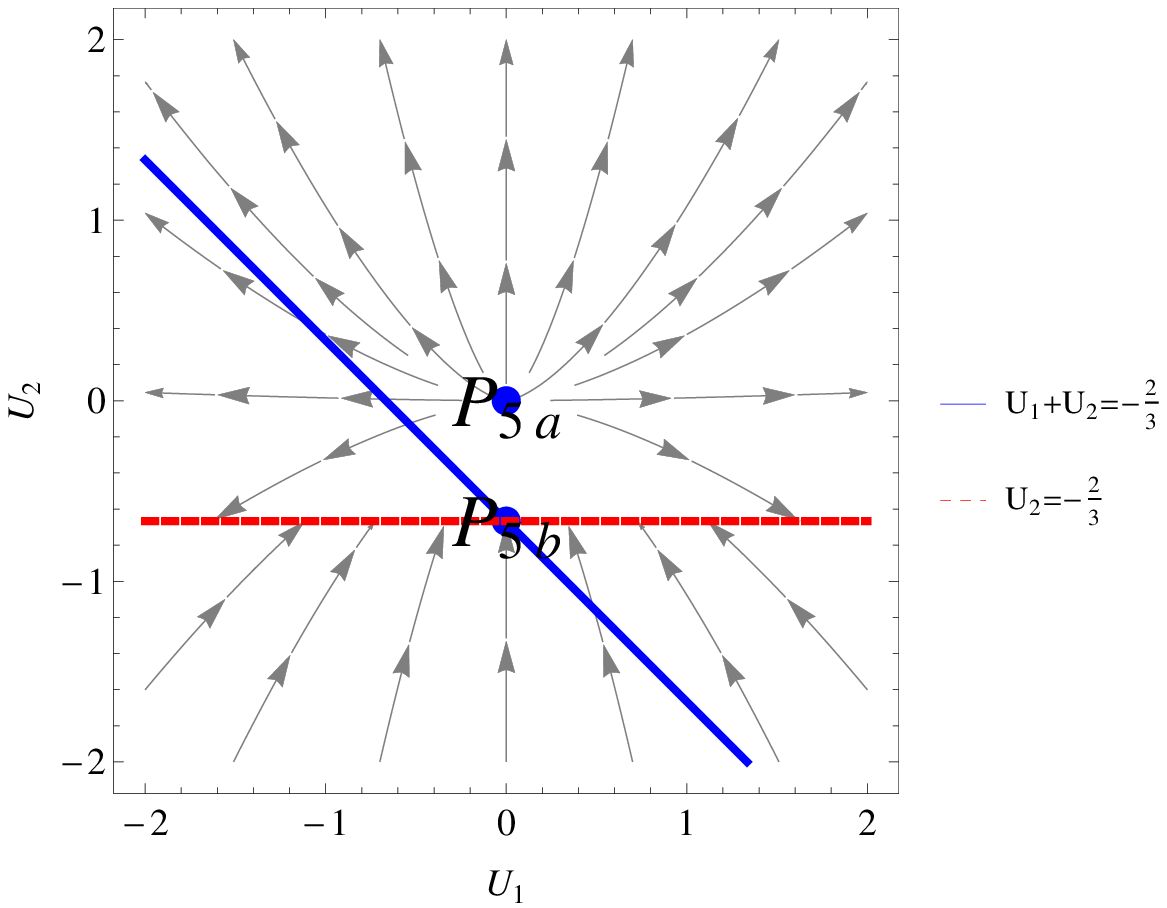}
\includegraphics[width=0.4\textwidth]{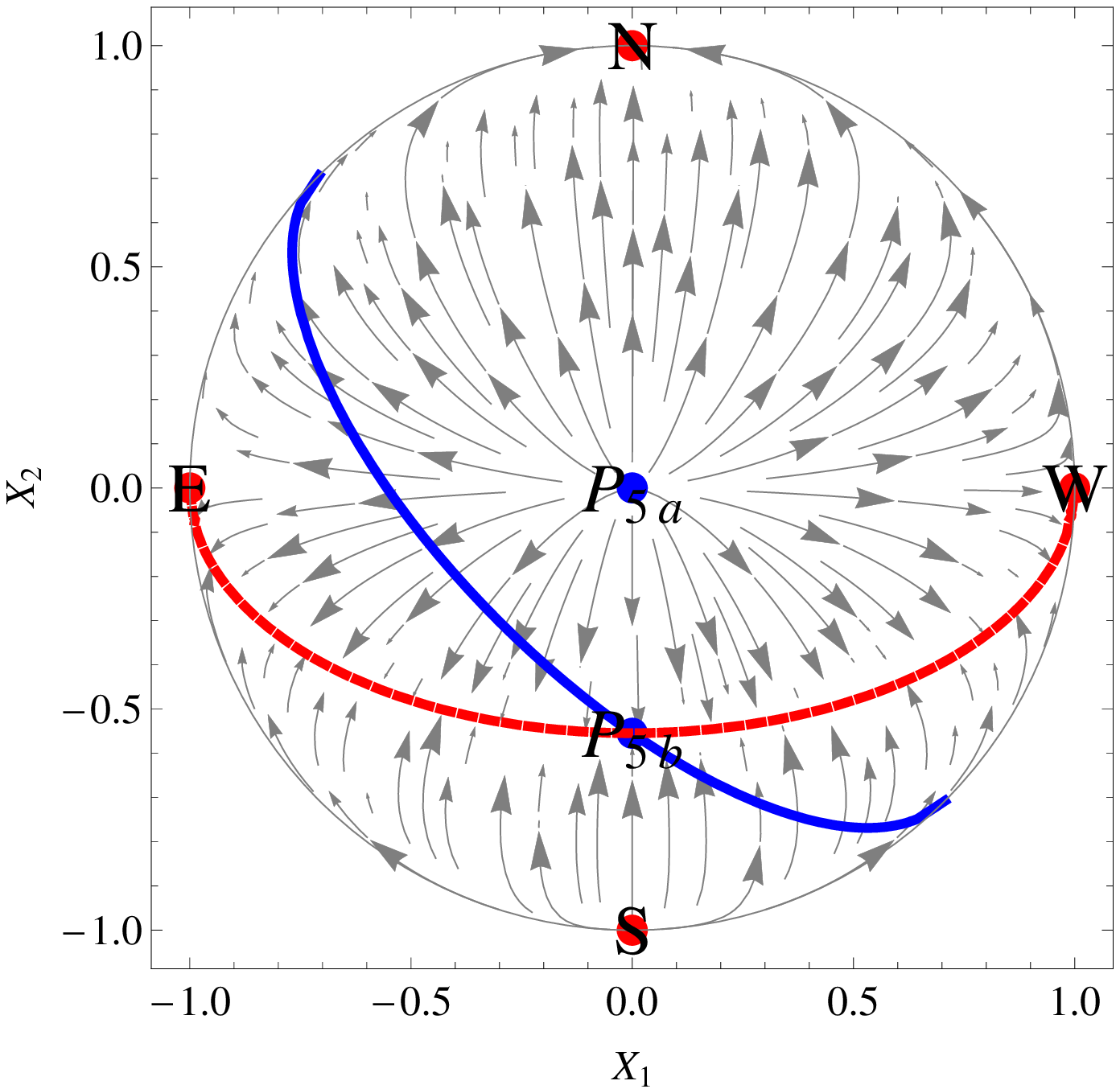}
\caption{\label{fig:DS5} Stream plot of the vector field defined by equations \eqref{eq4.39} and \eqref{eq4.40} for $x=\frac{2}{3}$, showing the dynamics at the invariant set $U_3=0$. The right panel shows the representation in Poincar\'e variables (see section \ref{fixpinf}). The region below the blue line $U_1+U_2=-\frac{2}{3}$ represents the physical portion of the phase space. The red dashed line represents the line of fixed points $P_{5b}$.}
\end{figure}

It follows from \eqref{eq4.39} and \eqref{eq4.40} that $P_{5b}$ is not longer a single fixed point in the case $x=\frac{2}{3}$, instead it becomes the line of fixed points $U_2=-\frac{2}{3}$ (represented by a red dashed line in figure \ref{fig:DS5}) and it is stable rather than saddle. This feature is not unexpected since we know that at bifurcation values of the parameters, as $x=\frac{2}{3}$, stability might change.

In the left panel of figure \ref{fig:DS5} it is presented the stream plot  of the vector field defined by equations \eqref{eq4.39} and \eqref{eq4.40} for $x=\frac{2}{3}$, which gives the dynamics at the invariant set $U_3=0$. The fixed point $P_{5a}$ is the past attractor but it lies outside of the physical portion of the phase space and the line $U_2=-\frac{2}{3}$ (represented by a red dashed line) is stable. In general this phase plane is unbounded, so that we have to search in the region at infinity. The right panel of figure  \ref{fig:DS5} shows the representation in Poincar\'e variables. The points at infinity are described in table \ref{tableInfinityU3zero}. For $x=\frac{2}{3}$ we have that the attractor at the infinite region of the phase space  is $N$, $W$ and $E$ are saddles and $S$ is the source.

\paragraph{Dynamics in the sector $\Gamma_-$.}
In this sector of the phase space $U_3<0$, so that the dynamics is restricted to
\begin{equation}
\left\{(U_1,U_2,U_3)\in\mathbb{R}^3:U_1+U_2\leq (x +1) U_3^2 -x,\quad -\frac{\sqrt{2}}{2}\leq U_3 \leq 0\right\}.
\end{equation}

We notice that the system of equations \eqref{eq3.37}--\eqref{eq3.39} can represent the dynamics of the sector $\Gamma_+$ as well as the dynamics of the sector $\Gamma_-$ under the time symmetry $\bar{\tau}\rightarrow -\bar{\tau}$, in such a way that the dynamics on both sectors is described as one to  have the time reversal behavior of the other.   That is, the dynamics in the invariant set $U_3=-\frac{\sqrt{2}}{2}$ (which contains the past attractor) corresponds to the time reversal of figures  \ref{fig:DS3} and \ref{fig:DS4}, meanwhile the dynamics in the invariant set $U_3=0$ (that contains the future attractor) corresponds to the time reversal of figures  \ref{fig:DS7} and \ref{fig:DS8}.

In section \ref{Sect:4.1} we have presented a detailed stability analysis of the scenario defined by equations  (\ref{Eq1})--(\ref{Eq5}) considering $f\neq0$ and pressureless matter. We have found that the acceleration (at early, late or intermediate epochs) is a generic feature of our model since the relevant fixed points are de Sitter solutions. In the next section we analyze the case of $f=0$. 
\subsection{Dynamical system analysis for $f=0$ with pressureless matter}
\label{SECT_4.2}

As we have commented in the section \ref{SecT:3_2}, the consistency equation \eqref{Eq_4_30} has to be satisfied leading to the following possibilities, $H=Z$ or $H=\pm\sqrt{Z_0^2-\frac{k}{a^2}}$.

\subsubsection{Case $H=Z$}

By defining as dimensionless variables,
\begin{equation}
\label{obs_H}
\Omega_{m}=\frac{\rho}{3 Z^2}, \quad  \Omega_{\beta}=\frac{\beta}{3 Z^2}, \quad \Omega_{k}=-\frac{k}{a^2 Z^2},
\end{equation}
we get from \eqref{EQ_4_34} the following constraint
\begin{equation}
\label{EQ_4.39}
\Omega_{m}+\Omega_{\beta }+\Omega_k=1,
\end{equation}
which leads to a compact phase space for $k=-1,0$, and to an unbounded phase space for $k=+1$.
By using a time variable such that $\Omega_i'=\frac{1}{Z} \dot \Omega_{i}$, from \eqref{LCDM} we obtain the following dynamical system 
\begin{align}
& \Omega_m'=\Omega_{m} (-2 \Omega_\beta +\Omega_{m}-1), \label{syst-case-1-a} \\ & \Omega_{\beta}'= \Omega_{\beta}  (-2 \Omega_{\beta} +\Omega_{m}+2), \label{syst-case-1-b}
\end{align}
where we have used \eqref{EQ_4.39} to replace $\Omega_k$. Because we are interested in the cosmological applications we have assumed $H>0$ in the previous analysis. 

Equations \eqref{syst-case-1-a}--\eqref{syst-case-1-b} are standard in cosmology, we summarize the results found elsewhere in table \ref{table4} (e.g. \cite{Joint0}), where the stability conditions of the critical points are shown. Additionally, we estimate the asymptotic behavior of the scale factor at each fixed point. We see that in the finite region of the phase space the future attractor corresponds to a de Sitter solution and the past attractor is the matter dominated universe. The fixed point $K_1$ corresponds to $\Omega_k=1$ and it only exists for $k=-1$. 

\begin{table}[t!]
\centering
\scalebox{0.9}{
\begin{tabular}{|c|c|c|c|c|}
\hline
Labels & $(\Omega_m, \Omega_\beta)$ & Eigenvalues& Stability & Asymptotic behavior of $a(t)$  \\\hline
$K_1$ & $(0,0)$ & $\{2, -1\}$ & saddle & $a=a_0 \pm \sqrt{t}$ (curvature dominated solution)\\\hline
$K_2$ & $(0,1)$ & $\{-3, -2\}$ & sink & $\sim a_0 e^{\sqrt{\frac{\beta}{3}} t}$ (de Sitter solution)\\\hline
$K_3$ & $(1,0)$ & $\{3, 1\}$ & source & $\sim t^{\frac{2}{3}}$ (matter dominated solution)\\\hline
\end{tabular}}
\caption{\label{table4} Stability of the critical points for the system \eqref{syst-case-1-a}--\eqref{syst-case-1-b}.}
\end{table}

\subsubsection{Case $Z^2+\frac{k}{a^2}= Z_0^2$}
In section \ref{SecT:3_2} we see that in this case it is possible to find an analytical solution. In this section we are interested in the evolution of the following dimensionless variables
\begin{equation}
\label{obs_2}
\bar{\Omega}_{m}=\frac{\rho}{3 N Z^2},\quad \bar{\Omega}_{\beta}=\frac{\beta}{3 Z^2},\quad \bar{\Omega}_{k}=-\frac{k}{a^2 Z^2},
\end{equation}
which satisfy the following constraint
\begin{equation}
\label{EQ_4.39b}
\bar{\Omega}_{m}+\bar{\Omega}_{\beta }+\bar{\Omega}_k=1.
\end{equation}
Assuming $0<Z_0^2<\beta$, we can define a new time variable in such a way that
\begin{equation}
\bar{\Omega}_i'=\frac{1}{Z}\left(\frac{Z_0^2}{\beta-Z_0^2}\right) \dot{\bar{\Omega}}_{i},
\end{equation}
then the dynamical system associated to $\bar{\Omega}_m$ and $\bar{\Omega}_{\beta}$ is given by
\begin{eqnarray}
\bar{\Omega}_m'&=&\frac{\bar{\Omega}_{m} \left(2 (\ell-1) \bar{\Omega}_\beta ^2+(\ell-1) \bar{\Omega}_\beta  (\bar{\Omega}_{m}+1)+\bar{\Omega}_{m} (-\ell (\bar{\Omega}_{m}-4)+\bar{\Omega}_{m}-1)\right)}{(\ell-1) (\bar{\Omega}_{m}-2 \bar{\Omega}_\beta )}, \label{syst-case-20}\\
\bar{\Omega}_{\beta}'&=&\bar{\Omega}_\beta  (1- \bar{\Omega}_\beta -\bar{\Omega}_{m}), \label{syst-case-2}
\end{eqnarray}
where \eqref{EQ_4.39b} was used to eliminate $\bar{\Omega}_k$, and $\ell= \frac{\beta}{3 Z_0^2}\geq \frac{1}{3}$.  

\begin{table}[h!]
\centering
\scalebox{0.9}{
\begin{tabular}{|c|c|c|c|c|}
\hline
Labels & $(\bar{\Omega}_{m}, \bar{\Omega}_\beta)$ & Eigenvalues& Stability & Asymptotic behavior of $a(t)$ \\\hline
$L_1$ & $(1-\ell, \ell)$ & $\{-1,\frac{3 \ell}{3 \ell-1}\}$  & saddle & $\sim a_0 e^{\frac{3 \ell -1}{2} \sqrt{\frac{\beta}{3\ell}} t}$\\\hline
$L_2$ & $(0,1)$ & $\{-\frac{3}{2},-1\}$ & sink &  $\sim a_0 e^{\sqrt{\frac{\beta}{3}} t}$ \\\hline
$L_3$ & $\left(\frac{1-4 \ell}{1-\ell}, 0\right)$& $\{-\frac{1-4 \ell}{1-\ell},\frac{3 \ell}{1-\ell}\}$  & source for $\frac{1}{3}<\ell<1$ & \\
 &&& sink for $\ell>1$&  $\sim a_0 e^{\frac{3 \ell -1}{2}\sqrt{\frac{\beta(1-\ell)}{3 \ell (1-4 \ell)}} t}$\\\hline
\end{tabular}}
\caption{\label{table5} Stability of the critical points of system \eqref{syst-case-20}--\eqref{syst-case-2}. }
\end{table}
In table \ref{table5} the stability conditions of the critical points of system \eqref{syst-case-20}--\eqref{syst-case-2} are summarized. We estimate the asymptotic behavior of the scale factor at each fixed point and we find that all the possible scenarios correspond to accelerated expansion.

\section{Final Remarks}\label{tres}
In this work we present a generalized Brans-Dicke theory in the framework of Horndeski theory including torsion, with the aim of analyze some cosmological consequences of the theory. {A remarkable feature of the studied model is that the source of the torsion are the non-minimal couplings between curvature and the scalar field, i.e., to assume a null torsion into the equations of motion (\ref{E_metrica})--(\ref{E_escalar}) necessarily freezes the dynamics of the scalar field.}

By considering pressureless matter and an effective torsional fluid in section \ref{cosmo} we were able to find conditions (\ref{conditions}) in order to get an analytical solution which describes several possible cosmological scenarios through the free parameters in the theory, see table \ref{table1}. Particularly, a cosmological scenario mimicking the dynamics of the late-time universe at background level was found.

In section \ref{cosmo} the most interesting scenario is the one where the expansion experiences a transition from decelerated to accelerated during evolution, and the torsion plays the role of an effective fluid with a time dependent state parameter. In figure \ref{fig5} we show many possible cosmological scenarios with a transition from decelerated to accelerated expansion, where the relevance of the effective torsional fluid could be dominant or sub-dominant in the past. Among all these scenarios, the more compatible with the standard cosmological scenario seems to be the one with the parameters fixed to $x=1.67,\ y=0.24$ and $z=0.56$, where the effective torsional fluid plays an important role today and in the future (where it asymptotically behaves as a cosmological constant, see figure \ref{figb}) but it becomes sub-dominant to the past where a pressureless matter component dominates at redshift around 200. An interesting characteristic of this scenario is that during the evolution the derivative of the Hubble expansion rate becomes positive (peculiarity non-observed in the standard cosmological scenario), which could represent a distinctive imprint of this kind of theory. Nevertheless, this last statement deserves more attention, given that scenarios such the ones in figure \ref{fig5} have been also described, for instance, in \cite{Algoner}, where the cosmological consequences of a scalar-tensor torsion-less scenario were presented. In this sense, besides to distinguish from the standard cosmological scenario our model must be distinguishable from a torsion-less scalar-tensor scenario.

The solution revised in section \ref{cosmo} is analytical and in this sense it corresponds to a particular case. In order to get more insight into solutions of equations \eqref{Eq1}--\eqref{Eq5} (with conditions \eqref{conditions}) we have performed a dynamical system analysis of the theory, where it is possible to get asymptotic behaviors without need of fixing initial conditions. We find that for $0<\beta<3 Z_e^2$, $Z_{e}>0$, the early and late-time attractors corresponds to de Sitter solutions which are accelerated, meanwhile, for $\beta <0$, the solutions are saddles such that the acceleration phase is a transient phenomena. Since almost all the fixed points have $q=-1$, the acceleration (at early, late or intermediate epochs) is a generic feature of our model, see table \ref{table2}.  
Furthermore, we have seen that no isolated fixed point at the finite region of the phase space corresponds to matter domination, nevertheless, the set $P_5$ in table \ref{table2} may contain a solution such that the torsional fluid mimics dust.

In the current article, for the sake of simplicity we used classical spin-less matter. A natural generalization would be to consider matter with a non-vanishing spin tensor playing the role of another torsion source, besides the non-minimal couplings with the scalar field. The full consistency of the structure could help to settle this issue. On the other hand, it is possible to add new terms to the lagrangian coupling torsion and the scalar field, this is natural in the context of conformal symmetry (work in progress) and it may help to close the gap between torsion-full and torsion-less dynamics in a more natural way than with the Lagrange multiplier shown in section \ref{uno}. The consequences of such procedure in a cosmological context remain to be seen.

\section*{Acknowledgments}
This work was partially funded by Comisi\'{o}n Nacional de Investigaci\'{o}n Cient\'{\i}fica y Tecnol\'{o}gica (CONICYT) with FONDECYT grants 1130653, 1150719 (FI). AC is partially supported by CONICYT through program Becas Chile de Postdoctorado en el Extranjero grant no. 74170121 and Direcci\'{o}n de Investigaci\'{o}n of Universidad del B\'{\i}o-B\'{\i}o through grants GI 150407/VC, 151307 3/R and GI-172309/C. PM and DN are funded by CONICYT scholarships from the Government of Chile through grants 21161574 and 21161099, respectively. GL thanks to Department of Mathematics at Universidad Cat\'olica del Norte for warm hospitality and financial support. We are grateful to Adolfo Toloza for many enlightening conversations.

\providecommand{\href}[2]{#2}\begingroup\raggedright\endgroup


\begin{thebibliography}{100}

\bibitem{libroFujiiMaeda} Y.~Fujii and K.~Maeda,
  \emph{The scalar-tensor theory of gravitation}, Cambridge University Press (2007).

\bibitem{Brans-Dicke}  C.~Brans and R.~H.~Dicke,
\emph{Mach's principle and a relativistic theory of gravitation}, \href{http://dx.doi.org/10.1103/PhysRev.124.925}{  \emph{Phys.\ Rev.\ } {\bf 124} (1961) 925.}

\bibitem{Horndeski}
G. W. Horndeski, \emph{Invariant Variational Principles and Field Theories}, PhD thesis, University of Waterloo (1973); G.~W.~Horndeski, \emph{Second-order scalar-tensor field equations in a four-dimensional space},
\href{http://dx.doi.org/10.1007/BF01807638}{ \emph{ Int.\ J.\ Theor.\ Phys.\ } {\bf 10} (1974) 363.}
  
\bibitem{dicke1968scalar} 
P.~J.~E.~Peebles and R.~H.~Dicke,  \emph{Origin of the Globular Star Clusters},
\href{http://dx.doi.org/10.1086/149811}{\emph{Astrophys.\ J.\ }  {\bf 154} (1968) 891.}
    
\bibitem{extended}D.~La and P.~J.~Steinhardt, \emph{Extended Inflationary Cosmology}, \href{http://dx.doi.org/10.1103/PhysRevLett.62.376}{\emph{Phys.\ Rev.\ Lett.\ } {\bf 62} (1989) 376. Erratum: [\emph{Phys.\ Rev.\ Lett.\  }{\bf 62} (1989) 1066].}

\bibitem{libroFaraoni}
V. Faraoni, \emph{Cosmology in Scalar-Tensor Gravity}, Springer (2004).

\bibitem{STDE} 
 B.~Boisseau, G.~Esposito-Farese, D.~Polarski and A.~A.~Starobinsky, \emph{Reconstruction of a scalar tensor theory of gravity in an accelerating universe},
\href{http://dx.doi.org/10.1103/PhysRevLett.85.2236}{\emph{Phys.\ Rev.\ Lett. }{\bf 85} (2000) 2236}
\href{https://arxiv.org/abs/gr-qc/0001066}{\tt [gr-qc/0001066]}; 
E.~Elizalde, S.~Nojiri and S.~D.~Odintsov,
\emph{Late-time cosmology in (phantom) scalar-tensor theory: Dark energy and the cosmic speed-up}
\href{http://dx.doi.org/10.1103/PhysRevD.70.043539}{\emph{Phys.\ Rev.\ D} {\bf 70} (2004) 043539}
\href{https://arxiv.org/abs/hep-th/0405034}{\tt [hep-th/0405034]};
S.~Capozziello, S.~Nojiri and S.~D.~Odintsov,
\emph{Dark energy: The Equation of state description versus scalar-tensor or modified gravity},
\href{http://dx.doi.org/10.1016/j.physletb.2006.01.065}{\emph{Phys.\ Lett.}\ B {\bf 634} (2006) 93}
\href{https://arxiv.org/abs/hep-th/0512118}{\tt [hep-th/0512118]};  
R.~Gannouji, D.~Polarski, A.~Ranquet and A.~A.~Starobinsky,
\emph{Scalar-Tensor Models of Normal and Phantom Dark Energy},
\href{http://dx.doi.org/10.1088/1475-7516/2006/09/016}{\emph{JCAP} {\bf 0609} (2006) 016}
\href{https://arxiv.org/abs/astro-ph/0606287}{[astro-ph/0606287]}. 
\bibitem{STDE2}
N.~Agarwal and R.~Bean,
\emph{The Dynamical viability of scalar-tensor gravity theories},
\href{http://dx.doi.org/10.1088/0264-9381/25/16/165001}{\emph{Class.\ Quant.\ Grav.\ }  {\bf 25} (2008) 165001}
  [\href{https://arxiv.org/abs/0708.3967}{{\tt arXiv:0708.3967 [astro-ph]}}]; A.~Cid, G.~Leon and Y.~Leyva,
\emph{Intermediate accelerated solutions as generic late-time attractors in a modified Jordan-Brans-Dicke theory},
\href{http://dx.doi.org/10.1088/1475-7516/2016/02/027}{\emph{JCAP} {\bf 1602} (2016) no.02,  027}
[\href{https://arxiv.org/abs/1506.00186}{\tt arXiv:1506.00186 [gr-qc]]}.
  
\bibitem{Cartan} 
E. Cartan, \emph{Sur une generalisation de la notion de courbure de Riemann et les espaces à torsion.} {\emph C. R. Acad. Sci. (Paris)}, \textbf{174} (1922) 593;
E.~Cartan, \emph{Sur les varietes a connexion affine et la theorie de la relativite generalisee. (premiere partie)},
\emph{  Annales Sci.\ Ecole Norm.\ Sup.\ } {\bf 40}, 325 (1923);  
E.~Cartan, \emph{Sur les varietes a connexion affine et la theorie de la relativite generalisee. (premiere partie) (Suite)},
\emph{Annales Sci.\ Ecole Norm.\ Sup.\ } {\bf 41}, 1 (1924).

\bibitem{Poplawski1} N.~J.~Poplawski,
\emph{Cosmology with torsion: An alternative to cosmic inflation},
\href{http://dx.doi.org/10.1016/j.physletb.2010.09.056, 10.1016/j.physletb.2011.05.047}{\emph{Phys.\ Lett.\ B} {\bf 694} (2010) 181. Erratum: [\emph{Phys.\ Lett.\ B} {\bf 701} (2011) 672]}
\href{https://arxiv.org/abs/1007.0587}{\tt  [arXiv:1007.0587 [astro-ph.CO]]}.

\bibitem{Poplawski2}  N.~J.~Poplawski,
 \emph{Nonsingular, big-bounce cosmology from spinor-torsion coupling},
\href{http://dx.doi.org/10.1103/PhysRevD.85.107502}{\emph{Phys.\ Rev.\ D} {\bf 85} (2012) 107502}
\href{https://arxiv.org/abs/1111.4595}{\tt [arXiv:1111.4595 [gr-qc]]}.

\bibitem{Zanelli-Toloza} 
A.~Toloza and J.~Zanelli,
\emph{Cosmology with scalar--Euler form coupling}, 
\href{http://dx.doi.org/10.1088/0264-9381/30/13/135003}{\emph{Class.\ Quant.\ Grav.} {\bf 30} (2013) 135003}  
\href{https://arxiv.org/abs/1301.0821}{\tt [arXiv:1301.0821 [gr-qc]]}.

\bibitem{nuestro-otro-paper} 
J.~Barrientos, F.~Cordonier-Tello, F.~Izaurieta, P.~Medina, D.~Narbona, E.~Rodr\'iguez and O.~Valdivia,
\emph{Nonminimal couplings, gravitational waves, and torsion in Horndeski’s theory},
\href{http://dx.doi.org/10.1103/PhysRevD.96.084023}{ Phys.\ Rev.\ D {\bf 96}, no. 8, 084023 (2017)}
\href{http://arxiv.org/abs/arXiv:1703.09686}{\tt [arXiv:1703.09686 [gr-qc]]}.

\bibitem{nonRiem}
D.~Puetzfeld,
\emph{Status of non-Riemannian cosmology},
\href{http://dx.doi.org/10.1016/j.newar.2005.01.022}{\emph{New Astron.\ Rev.}\  {\bf 49}, 59 (2005)}
\href{http://arxiv.org/abs/gr-qc/0404119}{[gr-qc/0404119]}.

\bibitem{KK}
O.~Castillo-Felisola, C.~Corral, S.~del Pino and F.~Ram\'irez,
\emph{Kaluza-Klein cosmology from five-dimensional Lovelock-Cartan theory},
\href{http://dx.doi.org/10.1103/PhysRevD.94.124020}{\emph{Phys.\ Rev.\ D} {\bf 94}, no. 12, 124020 (2016)}
\href{http://arxiv.org/abs/arXiv:1609.09045}{\tt [arXiv:1609.09045 [gr-qc]]}.

\bibitem{Pedro2}
S.~del Campo, R.~Herrera and P.~Labrana,
\emph{On the Stability of Jordan-Brans-Dicke Static Universe},
\href{http://dx.doi.org/10.1088/1475-7516/2009/07/006}{\emph{JCAP} {\bf 0907} (2009) 006}
\href{https://arxiv.org/abs/0905.0614}{\tt [arXiv:0905.0614 [gr-qc]]}.
  
\bibitem{Supergravity-VanProeyen} A. Van Proeyen, \emph{Supergravity}, Cambridge University Press (2012).

\bibitem{Karpathopoulos:2017arc} 
L.~Karpathopoulos, S.~Basilakos, G.~Leon, A.~Paliathanasis and M.~Tsamparlis, \emph{Cartan symmetries and global dynamical systems analysis in a higher-order modified teleparallel theory}
\href{http://arxiv.org/abs/arXiv:1709.02197}{\tt [arXiv:1709.02197 [gr-qc]]}.

\bibitem{Starobinsky} A.~A.~Starobinsky,
\emph{Disappearing cosmological constant in f(R) gravity},
\href{http://dx.doi.org/10.1134/S0021364007150027}{\emph{JETP Lett.}\ {\bf 86} (2007) 157}
\href{https://arxiv.org/abs/0706.2041}{\tt [arXiv:0706.2041 [astro-ph]]}.

\bibitem{wein} 
S. Weinberg, \emph{ Cosmology}, Oxford (2008).

\bibitem{Pedro}
P.~Labrana,
\emph{Emergent Universe Scenario and the Low CMB Multipoles},
\href{http://iopscience.iop.org/article/10.1088/1742-6596/720/1/012016}{\emph{J.\ Phys.\ Conf.\ Ser.} {\bf 720}, 012016 (2016)}; C. Rios, P. Labrana and A. Cid, \emph{The Emergent Universe and the Anomalies in the Cosmic Microwave Background}, \href{http://iopscience.iop.org/article/10.1088/1742-6596/720/1/012008/meta}{\emph{J.\ Phys.\ Conf.\ Ser.}  {\bf 720}, 012008 (2016)}.

\bibitem{Planck} 
P.~A.~R.~Ade {\it et al.} [Planck Collaboration],
\emph{Planck 2015 results. XIII. Cosmological parameters},
\href{tp://dx.doi.org/10.1051/0004-6361/201525830}{\emph{Astron.\ Astrophys.} {\bf 594} (2016) A13} \href{https://arxiv.org/abs/1502.01589}{\tt [arXiv:1502.01589 [astro-ph.CO]]}.
  
\bibitem{Algoner}
W.~C.~Algoner, H.~E.~S.~Velten and W.~Zimdahl,
\emph{Scalar-tensor extension of the $\Lambda$CDM model},
\href{http://dx.doi.org/10.1088/1475-7516/2016/11/034}{\emph{JCAP}\ {\bf 1611}, no. 11, 034 (2016)}
\href{http://arxiv.org/abs/arXiv:1607.03952}{\tt [arXiv:1607.03952 [gr-qc]]}.

\bibitem{Joint0}
J.~Wainwright and G.~F.~R. Ellis, \emph{Dynamical systems in cosmology}.
\newblock Cambridge University Press, 2005; A.~A. Coley, \emph{Dynamical systems and cosmology} in Astrophysics and Space Science Library Series, vol. 291, Springer, Germany (2003).

\bibitem{Joint1}
P.~G. Ferreira and M.~Joyce, \emph{{Structure formation with a selftuning scalar field}},
\href{http://dx.doi.org/10.1103/PhysRevLett.79.4740}{\emph{Phys. Rev. Lett.}
{\bf 79} (1997) 4740--4743},
\href{http://arxiv.org/abs/astro-ph/9707286}{{\tt [astro-ph/9707286]}};
E.~J. Copeland, A.~R. Liddle and D.~Wands, \emph{{Exponential potentials and cosmological scaling solutions}},
\href{http://dx.doi.org/10.1103/PhysRevD.57.4686}{\emph{Phys. Rev.} {\bf D57} (1998) 4686--4690}, \href{http://arxiv.org/abs/gr-qc/9711068}{{\tt [gr-qc/9711068]}}.
L.~Perko, \emph{Differential Equations and Dynamical Systems}.
\newblock Springer-Verlag GmbH, 2008;
X.-m. Chen, Y.-g. Gong and E.~N. Saridakis, \emph{{Phase-space analysis of interacting phantom cosmology}},
\href{http://dx.doi.org/10.1088/1475-7516/2009/04/001}{\emph{JCAP} {\bf 0904} (2009) 001}, \href{http://arxiv.org/abs/0812.1117}{{\tt [arXiv:0812.1117 [gr-qc]]}};
R.~Giambo and J.~Miritzis, \emph{{Energy exchange for homogeneous and isotropic universes with a scalar field coupled to matter}},
\href{http://dx.doi.org/10.1088/0264-9381/27/9/095003}{\emph{Class. Quant. Grav.} {\bf 27} (2010) 095003}, \href{http://arxiv.org/abs/0908.3452}{{\tt [arXiv:0908.3452 [gr-qc]]}};
D.~Escobar, C.~R. Fadragas, G.~Leon and Y.~Leyva, \emph{Phase space analysis of quintessence fields trapped in a Randall-Sundrum Braneworld: a refined study}, {\emph{Class. Quantum Grav.} {\bf 29} (2012) 175005},
\href{http://arxiv.org/abs/1110.1736}{{\tt [arXiv:1110.1736 [gr-qc]]}};
C.~Xu, E.~N. Saridakis and G.~Leon, \emph{{Phase-Space analysis of Teleparallel  Dark Energy}},
\href{http://dx.doi.org/10.1088/1475-7516/2012/07/005}{\emph{JCAP} {\bf 1207} (2012) 005} \href{http://arxiv.org/abs/1202.3781}{{\tt [arXiv:1202.3781 [gr-qc]]}};
G.~Leon and C.~Fadragas, \emph{Cosmological Dynamical Systems: And Their Applications}. LAP Lambert Academic Publishing, 2012 \href{http://arxiv.org/abs/1412.5701}{{\tt [arXiv:1412.5701 [gr-qc]]}};
D.~Escobar, C.~R. Fadragas, G.~Leon and Y.~Leyva, \emph{Phase space analysis of quintessence fields trapped in a randall-sundrum braneworld: anisotropic bianchi i brane with a positive dark radiation term}, {\emph{Class. Quantum Grav.} {\bf 29} (2012) 175006}
\href{http://arxiv.org/abs/1201.5672}{{\tt [arXiv:1201.5672 [gr-qc]]}};
D.~Escobar, C.~R. Fadragas, G.~Leon and Y.~Leyva, \emph{Asymptotic behavior of a scalar field with an arbitrary potential trapped on a Randall-Sundrum's Braneworld: the effect of a negative dark radiation term on a Bianchi I   brane}, {\emph{Astrophys. and Space Sci.} {\bf 349} (2014) 575} \href{http://arxiv.org/abs/1301.2570}{{\tt [arXiv:1301.2570 [gr-qc]]}};
C.~R. Fadragas, G.~Leon and E.~N. Saridakis, \emph{Dynamical analysis of anisotropic scalar-field cosmologies for a wide range of potentials}, {\emph{Class. Quant. Grav.} {\bf 31} (2014) 075018}
\href{http://arxiv.org/abs/1308.1658}{{\tt [arXiv:1308.1658 [gr-qc]]}};
S.~Cotsakis and G.~Kittou, \emph{{Flat limits of curved interacting cosmic fluids}}, \href{http://dx.doi.org/10.1103/PhysRevD.88.083514}{\emph{Phys. Rev.} {\bf D88} (2013) 083514} \href{http://arxiv.org/abs/1307.0377}{{\tt  [arXiv:1307.0377 [gr-qc]]}};
G.~Le\'on~Torres, \emph{{Qualitative analysis and characterization of two cosmologies including scalar fields}}.
\newblock PhD thesis, Marta Abreu Central U., Cuba, 2010
\href{http://arxiv.org/abs/1412.5665}{{\tt [arXiv:1412.5665 [gr-qc]]}};
C.~R. Fadragas and G.~Leon, \emph{{Some remarks about non-minimally coupled scalar field models}},
\href{http://dx.doi.org/10.1088/0264-9381/31/19/195011}{\emph{Class. Quant. Grav.} {\bf 31} (2014) 195011} \href{http://arxiv.org/abs/1405.2465}{{\tt
[arXiv:1405.2465 [gr-qc]]}};
S.~Wiggins, \emph{Introduction to Applied Nonlinear Dynamical Systems and Chaos}.\newblock Springer, 2010;
A.~Alho, J.~Hell and C.~Uggla, \emph{{Global dynamics and asymptotics for monomial scalar field potentials and perfect fluids}},
\href{http://dx.doi.org/10.1088/0264-9381/32/14/145005}{\emph{Class. Quant. Grav.} {\bf 32} (2015) 145005} \href{http://arxiv.org/abs/1503.06994}{{\tt
[arXiv:1503.06994 [gr-qc]] }}.
\end{thebibliography}
\end{document}